\documentclass[preprint]{aastex}
\input epsf

\newcommand\figenergy       {1}
\newcommand\figthreehalves  {2}
\newcommand\figfinalstate   {3}
\newcommand\figbigb         {4}
\newcommand\figrelease      {5}
\newcommand\figfoldscale    {6}
\newcommand\figfoldspectrum {7}
\newcommand\figbfcloss      {8}
\newcommand\figcji          {9}
\newcommand\figsubprandtla {10}
\newcommand\figsubprandtlb {11}
\newcommand\figbraginski   {12}
\newcommand\figalias       {13}

\newcommand\tabledyn      {2}

\newcommand\tablegalaxy   {4}
\newcommand\tablescales   {5}
\newcommand\tablepower    {6}
\newcommand\tablesima     {7}
\newcommand\tablealias    {8}

\newcommand\be{\begin{equation}}
\newcommand\ee{\end{equation}}
\newcommand\aaa{{\bf a}}
\newcommand\B{{\bf B}}
\newcommand\bb{{\bf b}}
\newcommand\f{{\bf f}}
\newcommand\kk{{\bf k}}

\newcommand\vv{{\bf v}}
\newcommand\x{{\bf x}}
\newcommand\zze{{\bf \zeta}}
\newcommand\bnabla{\mbox{\boldmath $\nabla$}}
\newcommand\hh{\hspace{1mm}}

\begin{document}

\nonumber
\setcounter{equation}{0}

\title{\mbox{The Nonlinear Turbulent Dynamo}}

\author{Jason Maron\altaffilmark{1} and Steve Cowley\altaffilmark{2}}
\affil{UCLA}

\altaffiltext{1}{maron@tapir.caltech.edu}

\altaffiltext{2}{steve.cowley@ic.ac.uk}

\begin{abstract} We simulate the evolution of an initially weak
magnetic field in forced turbulence for a range of Prandtl
numbers. The field grows exponentially with the Kulsrud-Anderson
$k^{3/2}$ spectrum until the magnetic energy approaches the
viscous-scale kinetic energy, where the magnetic forces then backreact
on the velocity. Further growth proceeds more slowly until a saturated
state is reached where the magnetic and kinetic energies are equal,
and where the magnetic energy exists primarily at the resistive
scale. We discuss the structure of this turbulence and the
extrapolation of the results to astrophysically-large Prandtl numbers.
\end{abstract}


\section{Introduction}

Microgauss magnetic fields are observed in spiral galaxies and between
galaxies in clusters [Zweibel \& Heiles (1997), Kronberg (1994),
Vallee (1998), Beck et. al. (1996)].  In cluster plasmas the fields
have coherence lengths of up to $10kpc$ (Taylor et. al., 1999).  In
galaxies the magnetic fields are ordered over the whole galaxy and
have energy densities comparable to the turbulent energy density.  It
is not known if the fields originated before during or after galaxy
formation.  Most current research centers around dynamo theories where
turbulent motions amplify the field from an initial weak seed field to
its present strength and structure [Ruzmaikin et. al. (1988), Kulsrud
(1999), Zweibel \& Heiles (1997), Beck et. al. (1996)] Indeed some
kind of dynamo seems to be the most plausible explanation of the
observations.  However, despite considerable progress over forty years
the dynamo theory is not complete and thus the history of galactic and
extra galactic fields is uncertain.  The galactic dynamo (if it
exists) differs from the more familiar solar dynamo [Mestel, Weiss]
and geodynamo [Glatzmaier \& Roberts 1995] in two key ways.  First the
disc geometry of the galaxy clearly affects the magnetic field
dynamics.  Secondly the ratio of viscosity to resistivity, the
magnetic Prandtl number ($P_r$), is of order $10^{15}$ in the warm
partially ionized interstellar medium but only $1 - 10^{-2}$ in the
solar convection zone.  We shall also discuss here the possibility of
dynamos in fully ionized plamas such as protogalaxies and early
galaxies.  In this paper we show that high Prandtl number dynamos are
profoundly different from low Prandtl number dynamos.

The galactic spatial scales and their associated timescales suggest a
{\em possible} scenario for field growth.  Elements of this scenario
are suggested by the work of Kulsrud and Anderson (1992) and,
particularly for Stage 3, Field, Blackman and Chou (1999).  We take
this scenario as a framework for discussion and not as proven.  Indeed
our calculations already expose flaws in the scenario.  First consider
the important space and time scales from the largest to the smallest.
The galaxy itself is about $10^{10}$ years old, $10-15kpc
(=\lambda_g)$ across and rotates with velocity $\sim 200km/s$ once
every $2\times 10^8$ years.  Supernovae produce random velocities of
order $10km/s$ on a scale $100pc (= \lambda_f)$ and with timescales of
order $10^{7}$ years.  Approximately $5-15\% $ of the kinetic energy
at the supernova scale is helical [Moffatt 1978]. Without a magnetic
field the kinetic energy cascades to small viscous scales,
$\lambda_{\nu}$, where $\lambda_{\nu}\sim 0.1 - 0.01pc$.  The viscous
eddies turnover on a timescale $\tau_\nu \sim 10^5$ years.  Finally
the smallest scale is the resistive scale, $\lambda_{\eta}$ which is
typically $10^3km$ -- truly negligibly small.  We will assume that the
initial field is very weak.  Such fields can be made in many different
ways, see for instance Gnedin \& Ferrara, \& Zweibel (2000) who
find that fields of order $10^{-18}$ Gauss can be made in shocks
during the re-ionization phase of structure formation.  We divide the
growth into three stages.

{\bf Stage 1. The Kinematic Small Scale Field Dynamo.}  During this
stage the field is too weak to affect the velocity at any scale.
Since the eddies at the viscous scale turnover the fastest they
amplify the field first.  Therefore the growth time is approximately
$10^5$ years in our galaxy.  This growth was first predicted by
Batchelor (1950), and later Kazantsev (1968), who also investigated
the structure of the field, and still more recently Kulsrud and
Anderson gave a spectral theory of the process (1992).  In the 1990s
the kinematic theory was considerably refined [Schekochihin,
et. al. (2001), Chertkov et. al. (1999)] and it is well understood.
Much of what is understood comes from the short correlation time
kinematic model in which the velocity correlation time is assumed
infinitesimal.  The important features of the magnetic field evolution
in this model are: first, the magnetic spectrum $E_b(k,t)$ rises as
$k^{{3/2}}$ at all $k$ much less than the peak, $k_{p}$.  Second, for
$k<k_{p}$, $E_b(k,t)$ grows as $\exp{(3/4\gamma t)}$ at fixed $k$ with
$\gamma$ roughly the turnover rate of the viscous scale eddies.  The
peak wavenumber, $k_{p}$ increases as $k_{p}\propto \exp{(3/5\gamma
t)}$ until it reaches, and remains at, the resistive scale
($k_{\eta}\sim 1/(\lambda_{\eta}) \sim \sqrt{\gamma / \eta}$ with
$\eta$ the resistivity).  Finally the magnetic field is in a folded
state [Schekochihin et. al., 2001], where the variation of $\bf B$
along itself ($<|{\bf B}\cdot \nabla {\bf B}|^{2}>$) is much smaller
than the variation of B across itself ($<|{\bf B}\times (\nabla
\times{\bf B})|^{2}>$).  All these features of the kinematic phase
evolution are seen, at least qualitatively, in the early stages of our
simulations.  Clearly the field predicted by the small scale field
dynamo is on scales much smaller than the observed galactic field.
Therefore Stage 1 is necessarily a transitory stage.  When the
magnetic field energy becomes of order the energy in the viscous scale
eddies the kinematic stage (Stage 1) ends.  In the galaxy this
corresponds to a magnetic field strength of $0.1\mu G$.

{\bf Stage 2. Approach to Equipartition.}  While magnetic forces at
the end of Stage 1 can change the viscous scale flows they are not
strong enough to affect the more energetic larger scale motions.  The
strain of these larger scale motions continues to amplify the field.
One would expect that they act on the field in a way similar to the
viscous scale eddies in Stage 1, but on a slower timescale.  As the
field grows more of the velocity spectrum is affected by the magnetic
forces (see Section 5.).  Eventually on a timescale of a few large
eddy turnover times the magnetic energy becomes of order the kinetic
energy of the large stirring scale motions.  In the galaxy the
timescale for this is the supernova stirring time -- roughly $10^{7}$
years.  On this timescale the galactic rotation and the helical
component of the turbulence have negligible effect.

This stage is porrly understood and it is the main subject of this
paper.  In 1981 Meneguzzi and Poquet computed the turbulent
amplification of magnetic fields in a $P_r\sim 1$ plasma.  They showed
that without helicity the magnetic energy grew until it saturated at
about $15\% $ of the kinetic energy.  The magnetic and kinetic spectra
resemble our higher resolution, $P_r\sim 1$, computations (for example
see Fig. \figfinalstate). There is no evidence from these simulations
(or our own) that there is energy equipartition scale by scale. In the
$P_r\gg 1$ case, the majority of the magnetic energy is in the scales
between the viscous and (smaller) resistive scale.  Almost none of the
kinetic energy is contained in these sub-viscous scales.

Recently there has been considerable interest in simulating Alfven
wave turbulence [Maron \& Goldreich (2001), Cho \& Vishniac (2000),
Biskamp \& Muller (2000)].  Identical magnetic and kinetic power law
spectra ($k^{-\alpha}$ with $\alpha = 1.5 - 1.6$) are seen. These
simulations start with magnetic energy on the large scale that is
stronger or of order the kinetic energy of the forcing scale.  Dynamo
simulations must, of course, start with small fields.  It is not clear
that the final state is independant of the initial conditions -- at
least for times less than the resistive time of the large scales.

At the end of Stage 2. the magnetic energy is expected to be of order
the kinetic energy of the forcing motions.  The observed field is on
scales bigger than the supernova forcing scale.  There is no evidence
from our simulations that there is any large scale field at the end of
Stage 2 -- indeed the majority of the magnetic energy is in the
subviscous scales.

{\bf Stage 3. Large Scale Field Growth.}

The final stage of the magnetic field evolution is growth of the field
on the largest scale, the galactic scale.  It is believed [Poquet
et. al. (1977), Meneguzzi \& Poquet (1981), Field et. al. (1999),
Brandenburg (2000)] that the helicity of the turbulence plays a key
role in this {\em inverse cascade}.  Indeed Field, Blackman and Chou
(1999) have argued that this stage is very similar to the kinematic
($\alpha -\omega$) mean field dynamo [Parker (1979), Moffatt (1978),
Ruzmaikin (1988)].  The estimated timescale for the mean field dynamo
in the galaxy is $2\times 10^{8}$ years [Field et. al., (1999),
Kulsrud (1999)].  This timescale is controlled by the slow rotation of
the galaxy that gives $\omega$ and drives the helicity in the
turbulence to give $\alpha$.  The fraction of the turbulent energy in
the galaxy that is helical is about $5-15\%$??  Recent simulations by
Maron and Blackman (2001) suggest that large scale field
growth at such low helical fractions is not possible (although it is
at $100\%$ helicity [Brandenburg (2000)].  They also show that the
fractional helicity needed to see large scale field growth increases
with magnetic Prandtl number.  There is an ongoing debate about the
effect of small scale fields on the mean field dynamo -- several
authors have claimed a quenching (suppression) of the alpha effect
[Hughes \& Cattaneo (1996), Gruzinov et. al. (1996), Bhattachargee
(1996)].  Blackman and Field (2000) have pointed out the important
role of boundary conditions that allow the magnetic helicity of one
sign to be discarded.  Without such boundary conditions they claim
that mean field growth is on a negligably slow resistive time
[Gilbert].  The field we obtain at the end of stage 2 is highly
striped (i.e. it reverses in sign on a small resistive scale).  Since
the mean field dynamo motions will amplify both the positive and
negative stripes one would expect amplification of the small scale
field to dominate.  However we will not consider this stage in any
detail in this paper.  It is clear that this stage is dependant on the
geometry and physical conditions in the galaxy and therefore
simulations in simplified homogeneous models may be missleading.


There are four key results from our simulations.  First, the energy of
the magnetic field grows to a saturated level equal to the kinetic
energy of the stirred flow.  Second, the magnetic field energy is
contained in the small resistive scales. Third, the field is in the
form of long thin folds (stripes) where the variation of ${\bf B}$
across itself is much faster than the variation of ${\bf B}$ along
itself.  Fourth, the predominant straining and folding of the field in
saturation comes from the stirring scale motions.

\section{Equations}

We investigate MHD turbulence in situations of interest to the origin
of the galactic magnetic field. Various regimes exist depending on the
state of the dynamical medium, specifically on whether it is charged
or neutral.  In an ionized plasma, if the magnetic field is
sufficiently strong, charged particle transport is confined to
magnetic fieldlines and the kinematic diffusivity has the anisotropic
Braginskii form (section \ref{sec:braginsky}).  If neutrals are
present, viscosity is isotropic. If additionally the magnetic energy
density is sufficiently high, ambipolar diffusion can arise from the
differential ion-neutral motion.
Magnetic diffusivity in all cases is due to resistivity.  We also
consider the artificial but illustrative problem involving ordinary
viscosity and resistivity (section \ref{sec:pure}). We consider these
regimes in the context of the galaxy and protogalaxy, where kinematic
diffusivities greatly exceed magnetic diffusivities.


The equations of magnetohydrodynamics (MHD) are
\begin{equation} \label{eq:momentum}
\rho\left(\partial_t \vv + \vv \cdot \bnabla \vv\right)
= - \bnabla\left(P + \frac{\B^2}{8 \pi}\right) - \bnabla \cdot \Pi
+ \frac{1}{4 \pi} \B \cdot \bnabla \B + \rho \nu \nabla^2 \vv
\end{equation}
\begin{equation} \label{eq:induction}
\partial_t \B = \bnabla \times ( \vv \times \B) + \eta \nabla^2 \B
\end{equation}
\begin{equation} \label{eq:continuity}
\partial_t \rho+\bnabla \cdot (\rho\vv) = 0,
\end{equation}
\begin{equation} \label{eq:passive}
\partial_t c + \bnabla \cdot (c \vv) = \nu_c \nabla^2 c
\end{equation}
\begin{equation} \label{eq:divB}
\bnabla \cdot \B = 0
\end{equation}
\begin{equation} \label{eq:braginski}
\Pi = - 3 \nu_{ii} (\hat{B}\hat{B} - \frac{1}{3}I)
(\hat{B}\hat{B} - \frac{1}{3}I) : \bnabla \vv
\end{equation}
\vspace{0.2 cm}



\begin{center}
\begin{tabular}{llll}
\hline
$\vv$     & fluid velocity        & $\B$  & magnetic field            \\
$\rho$    & fluid density         & $P$    & fluid pressure            \\
$\nu$     & momentum diffusivity  & $\eta  $& magnetic diffusivity      \\
$c$       & passive scalar density& $\nu_c$& passive scalar diffusivity\\
$\Pi$     & Braginskii pressure    & $\hat{B}$& Magnetic unit vector \\
\hline
\end{tabular} \end{center}

\begin{center} {\it Table \tabledyn: Dynamical variables} \end{center}


\noindent $\Pi$ is the Braginskii tensor, which applies when charged
particle transport is confined to fieldlines. Ambipolar diffusion can
optionally be added if the fluid is only partially ionized, and if
magnetic forces are strong enough to generate significant differential
ion-neutral velocities.


We simplify equations \ref{eq:momentum}-\ref{eq:passive} for
applications in this paper. The magnetic field is measured in velocity
units with $\bb \equiv \B/ \sqrt{4 \pi}$. Incompressibility is assumed
throughout, so we set $\rho = 1$. We these modifications, equations
\ref{eq:momentum}-\ref{eq:passive} transform to

\begin{equation}
\partial_t \vv = - \vv \cdot \bnabla \vv- \bnabla (P + \frac{\bb^2}{2})
- \bnabla \cdot \Pi + \bb \cdot \bnabla \bb + \nu \nabla^{2} \vv,
\label{eq:mhda}
\end{equation}
\begin{equation}
\partial_t\bb = \bnabla \times ( \vv \times \bb )
+ \eta \nabla^{2}\bb,
\label{eq:mhdb}
\end{equation}
\begin{equation}
\bnabla \cdot \vv = 0, \hspace{12mm} \bnabla \cdot \bb = 0.
\label{eq:mhdc}
\end{equation}
\be
\partial_t c+\vv\cdot\bnabla c=\nu_c \nabla^{2} c.
\ee


Incompressibility is implemented by first calculating $\partial_t \vv$
without the fluid and magnetic pressure terms, and then by projecting
the resulting Fourier components transverse to $\kk$. This procedure
accounts for the divergence introduced by the Braginskii term.



\subsection{Tensor viscosity} \label{sec:braginsky}

Particle transport in a fully ionized medium is confined to fieldlines
if the collision length exceeds the cyclotron radius:
$$ B \hh
\stackrel{>}{\sim} \hh \frac{c \hh m_p^{1/2} n_p e^3} {k^{3/2}}
T^{-3/2} \\ \sim \hh 2.7 \cdot 10^{-6} n_p T^{3/2} $$
This condition
is easily satisfied except possibly in the early protogalaxy where the
field may have been very weak. For ionized plasmas, we use the
Braginskii pressure (Eq. \ref{eq:braginski}) in place of the Laplacian
viscosity.

\section{Simulation}

\subsection{Scales} \label{Simulation} \label{scales}

We state our assumptions for the parameters of galactic turbulence in
table \tablegalaxy, followed by a discussion of our inability to
directly simulate them.

\begin{center} {\it Table \tablegalaxy: Galactic turbulence parameters}
\end{center}

\begin{center}
\begin{tabular}{llll}
\hline
Quantity          &Symbol           & Galactic & Protogalactic \\
\hline
Magnetic coherence scale&$\lambda_g$&$10^{22}$ cm   & - \\
Forcing scale     &$\lambda_f$    &$10^{21}$ cm    &$10^{23}$ cm \\
Forcing velocity  &$v_f$          &$10^{6}$ cm/s   &$10^{7}$ cm/s \\
Viscous scale     &$\lambda_\nu$  &$10^{17}$ cm    &$10^{22}$ cm\\
Resistive scale   &$\lambda_\eta$ &$10^{12}$ cm    &$10^{12}$ cm\\
Viscosity         &$\nu$          &$10^{20}$ ${\rm cm}^2/s$ &$10^{29}$
                                                              ${\rm cm}^2/s$ \\
Magnetic diffusivity&$\eta$       &$10^{14}$ ${\rm cm}^2/s$ &$10^{10}$
                                                              ${\rm cm}^2/s$ \\
Prandtl number    &$P$            &$10^6$          &$10^{19}$\\
Neutral free path &$\lambda_{nn}$ &$10^{15}$ cm    & - \\
Ion free path     &$\lambda_{ii}$ &$10^{13}$ cm    &$10^{22}$ cm \\
Temperature       &$T$            &$10^4$ K        &$10^7$ K \\
Proton density    &$n$            &1${\rm cm}^{-3}$&$10^{-2}{\rm cm}^{-3}$ \\
\hline
\end{tabular}
\end{center}

The range of scales involved in the galactic dynamo poses the chief
obstacle to simulation. A spectral grid of $256^3$ elements delivers a
scale range of $\sim 50$, which is enough to capture self-similar
dynamics or to study transitions between regimes occuring at different
scales.  The phases of the dynamo can be simulated separately and then
spliced together to construct a complete picture. Limitations exist of
course, most notably when we consider the growth of a large scale
field in the presence of a tangled small scale field. In the galactic
setting, these scales differ by orders of magnitude.

In the high-Prandtl linear regime, the dominant magnetic structure
exists between the viscous and the resistive scales. We find that at
least an order of magnitude of separation between them is necessary to
capture the $k^{3/2}$ Kulsrud-Anderson spectrum. In our simulations, a
Prandtl number of at least $2500$ is required.

Our principal concern regarding the nonlinear stage is the final state
of the magnetic field in the limit of large inertial range and Prandtl
number, and how long it takes to get there. Specifically, is this
field dominated by large or small scale structure, or equivalently how
does the spectral index compare to $-1$?  Finally, does the result for
larger Prandtl number differ from that for $P=1$?

In the nonlinear stage, the magnetic field is strong enough to
backreact on the turbulence. This first occurs at the viscous scale,
then at successively larger scales in the inertial range as the
magnetic field grows.  The objective is to study the backreaction on
the inertial-range structure when the magnetic energy is at the
resistive scale. It is important that the forcing occur at a
substantially larger scale than the viscosity so that it doesn't
interfere with the magnetic backreaction. Thus we require
$\lambda_\eta << \lambda_\nu << \lambda_f$. In a $256^3$ spectral
simulation, the effective scale range is $\sim 50$, and so we can have
a factor of $\sim 7$ separating each scale. To study the role of
viscosity and the inertial range, we ran 5 simulations with
$\lambda_\nu$ engineered to have a sequence of values from
$\lambda_\eta$ to $\lambda_f$.

\subsection{Computational scales} \label{computationalscales}

Table \tablescales \, defines the scales that occur in the simulation
of the turbulent dynamo. When the Prandtl number is greater than or
equal to unity, the scales have the ordering $\lambda_f > \lambda_\nu
\geq \lambda_\eta$. The dynamics occur between $\lambda_f$ and
$\lambda_a$, and so we set $\lambda_f \sim 1$ and $\lambda_\eta \sim
\lambda_a$, where the simulation volume is a cube of side length $1$
and $\lambda_a$ is the smallest resolved scale.  We ran a sequence of
simulations with $\lambda_\nu$ varying between $\lambda_f$ and
$\lambda_a$.  A forcing power of unity at the outer scale maintains an
RMS outer scale velocity of $v_f \sim 1$.

\begin{center} {\it Table \tablescales: Computational scales} \end{center}
\begin{center}
\begin{tabular}{lllll}
\hline
$v_\lambda$&Velocity at scale $\lambda$&\hspace{1cm}&$b_\lambda$&
Magnetic field at scale $\lambda$\\
$\nu$ &Kinematic diffusivity &\hspace{1cm}& $\eta$& Resistive diffusivity\\
$\lambda_\nu$&Viscous scale&\hspace{1cm}&$\lambda_\eta$& Resistive scale\\
$\lambda_f$& Forcing scale ($=1$)&&$\vv_f$&Outer scale RMS velocity (=1)\\
$\lambda_s$&Shear scale&\hspace{1cm}&$v_s$&Shear velocity\\
$\lambda_a=3/N$&Aliasing (resolution) scale&\hspace{1cm}&$N^3$& Grid size\\
$t_f \sim \lambda_f/v_f$&Forcing timescale&\hspace{1cm}&
                         $t_s \sim \lambda_s/v_s$&Shear timescale\\
$\lambda_\perp$&Transverse folding scale&\hspace{1cm}&$\lambda_\parallel$&
Longitudinal folding scale\\
$f=\lambda_\parallel/\lambda_\perp$&Fieldline folding factor&\hspace{1cm}&
$s=k/(2\pi)=1/\lambda$& Wavenumber\\
\hline
\end{tabular} \vspace{0.2 cm}
\end{center}

During the weak magnetic field regime, the following scalings enable
us to set $\lambda_\nu$ and $\lambda_\eta$ by varying $\nu$ and
$\eta$.  Below the forcing scale and above the viscous scale, the
velocity has the form of a Kolmogorov cascade with $v_\lambda \sim v_f
(\lambda/\lambda_f)^{1/3}$. The eddy time is $t \sim \lambda /
v_\lambda$. The viscous scale arises from equating the eddy time with
the viscous time.

\begin{equation}
\lambda_\nu \sim (\nu \lambda_f^{1/3} / v_f)^{3/4}
\end{equation}

Below the viscous scale, the velocity has the form of a uniform shear
with the same timescale as the eddy time at the viscous scale: $t_\nu
\sim \lambda_f^{1/2} \nu^{1/2} v_f^{-3/2}$.  Equating this with the
resistive time, $t_\eta \sim \lambda_\eta^2/\eta$, we obtain the
resistive scale $\lambda_\eta \sim \eta^{1/2} \lambda_f^{1/4}
\nu^{1/4} v_f^{-3/4}$.  We engineer $\eta$ so that $\lambda_\eta$ is
equal to the grid scale $\lambda_a \sim \lambda_f / N$, where $N$ is
the number of grid elements on each cube edge.  This establishes a
condition on $\eta$ as a function of $\nu$:

\begin{equation}
\eta_g \sim \frac{\lambda_f^{3/2} v_f^{3/2}}{\nu^{1/2} N^2}
\end{equation}

\noindent Equating the viscous and resistive times, the Prandtl number is
related to the ratio of scales as \begin{equation}
\frac{\lambda_\nu}{\lambda_\eta} \sim \frac{\nu^{1/2}}{\eta^{1/2}}
\sim P_r^{1/2} \end{equation} In designing a simulation with a
specified Prandtl number greater than or equal to unity, one uses the
above scalings to select $\nu$ and $\eta$.  For studies of the
kinematic backreaction, it is desirable to have $\lambda_\nu$ small
enough so that a true kinematic inertial range exists, say
$\lambda_\nu \sim \lambda_f/8$.

Once the magnetic field is strong enough to backreact on the velocity,
the shear scale is set by the eddies which have the same energy
density as the magnetic field. In the saturated state, the magnetic
energy grows to the forcing energy, and the shear time increases to
the forcing time.  Simultaneously, the resistive scale increases by a
factor of $(\lambda_f/\lambda_\nu)^{1/3}$ from its value in the linear
regime.

\subsection{Code}

We withhold for now all details about the code not necessary for
understanding the results, which we put off until section \ref{code}.
The equations of MHD are solved spectrally and incompressibly, and
with periodic boundaries. A standard $256^3$ grid with spatial
dimension $1^3$ has Fourier wavenumbers $s = k / (2\pi)$ extending
from -85 to +85, where we have employed the 2/3 aliasing
truncation. Wavenumbers and physical scales are related by $\lambda k
= \lambda 2 \pi s = 2 \pi$. Viscosity and resistivity are of the $k^2$
type ($\nu \nabla^2 v$ and $\eta \nabla^2 b$) unless the Braginskii
viscosity is invoked.  Finally, define the one-dimensional kinetic and
magnetic energy spectra as \begin{equation} E_v = \int E_v(s) d\!s
\hspace{15mm} E_b = \int E_b(s) d\!s. \end{equation} The parameters
for each simulation are given in table \tablesima.

\subsection{Uniform field and helicity}

The simulations in this work have zero mean magnetic field and are
forced with zero mean kinetic helicity.  Fractional fluctuations in
the kinetic helicity exist at the level of 10 percent. The magnetic
helicity is initially zero and subsequently fluctuates about zero at
an amplitude of 5 percent of the maximum potential magnetic helicity.

Our simulations have no mean magnetic field or helicity, to establish
what large-scale can be created in their absence. This is the
nonlinear kinematic dynamo problem. If indeed the final state is
dominated by small-scale magnetic energy, then the necessity of other
mechanisms such as helicity or Keplerian shear may be invoked for the
next stage of the dynamo.

\subsection{Folded fieldline structure} \label{structure}

The magnetic field, in the course of amplification and entanglement, develops
structure not identifiable in the power spectrum. Define

\begin{equation} \label{eq:sa}
k_\perp^2 = <(\bb \times \nabla \times \bb)^2> / <b^4> \hspace{10mm}
k_\parallel^2 = <(\bb \cdot \nabla \bb)^2> / <b^4>
\end{equation}
\begin{equation} \label{eq:sb}
k_P^2 = <\left( \nabla \frac{b^2}{2} \right)^2> / <b^4> \hspace{10mm}
k_\circ^2 = <(\bb \cdot \nabla \times \bb)^2> / <b^4>
\end{equation}

Define a fieldline folding factor $f = k_\perp/k_\parallel$,
which parameterizes the magnetic tension per energy and the unwinding
time of fieldlines. Also define a measure of the magnetic scale based on the
power spectrum:

\begin{equation}
k_b^2 = \int k^2 E_b(k) dk / \int E_b(k) dk
\end{equation}

We normalize with $<b^4>$ instead of $<b^2>^2$ because it has the same
kurtosis statistics as the squared magnetic force terms.  In the
simulations, $k_\perp/k_b \sim 0.58$ with fluctuations of 3 percent,
whereas the kurtosis varies by 200 percent (Schekochihin, et. al.,
2001).  Therefore, either $k_\perp$ or $k_b$ may be used to define the
magnetic scale.


\subsection{Timescales} We identify timescales for resistivity
($t_\eta$), vorticity ($t_w$), and shear ($t_s$) by defining:

\begin{equation}
t_\eta = \frac{1}{k_\perp^2 \eta} \hspace{12mm}
t_{w} = 2 \pi <(\nabla \times v)^2>^{-1/2} \hspace{12mm}
t_s = <b^2>
\end{equation}

$t_\eta$ reflects a balance between shear and resistivity.  $t_w$
corresponds to the eddy shear time and the magnetic growth rate during
the linear regime.  $t_s$ is indirectly linked to the shear time
through $t_s \sim b^2 \sim v_s^2 \sim \lambda_s/v_s$, where $b$ is the
saturation magnetic field, and $\lambda_s$ and $v_s$ correspond to the
scale and velocity of the dominant shear.

\subsection{Viscous scale}

We define the viscous scale $\lambda_\nu$ to correspond to the peak of
the function $k^3 E_v(k)$ during the linear regime, which reflects the
dominant scale of the shear.

\subsection{Spandex waves} \label{spandex}

MHD interactions exist which are non-local in $k$ space. For instance,
large-scale shear can transfer energy directly to small-scale magnetic
fields. Alfv\'en waves are an oscillatory phenomenon where small-scale
perturbations interact with a uniform magnetic field. A similar
oscillatory phenomenon exists in a tangled magnetic field. Here, a
large-scale velocity perturbation generates a backreaction in a
small-scale tangled field. Locally, magnetic forces act in all
directions, but spatially averaged, they tend to act collectively to
oppose the original large-scale perturbation.

Define a Lagrangian displacement field $\zeta = \hat{\aaa}
e^{i(\kk\cdot\x-\omega t)}$, with $\kk \cdot \hat{\aaa} = 0$ and $\vv
= d_t \zeta$.  Let $\bb = \bb_0 + \hat{\aaa} (\bb_0\cdot\kk) i
e^{i(\kk\cdot\x-\omega t)}$, where $\bb_0$ is the static
non-oscillatory part with an assumed scale of much less than
$k^{-1}$.  Any other time evolution in $\vv$ and $\bb$ is neglected,
as well as viscosity and resistivity. $\bb$ satisfies the induction
equation $\bb = \int d_t \bb \, d\!t = \bb \cdot \nabla
\zze$. Returing to the Navier Stokes equation, $<d_t \vv> = - \omega^2
e^{i(\kk\cdot\x - \omega t)} = <\bb \cdot \nabla \bb> = <\bb_0 \cdot
\nabla \bb_0> - <(\bb_0 \cdot \kk)^2> e^{i(\kk \cdot x - \omega
t)}$. An average is taken over scale $k$, which implies $<\bb_0 \cdot
\nabla \bb_0> \sim 0$ and $<(\bb_0 \cdot \kk)^2> \sim <(\bb_0)^2> k^2
/ 3$. $\zze$ has an oscillatory eigenmode with phase speed
$<(\bb_0)^2/3>^{1/2}$. This is equivalent to the Alfv\'en speed if we
only consider the averaged magnetic field component along $\kk$.

\section{Exponential growth of a weak magnetic field}
\label{sec:pure}

The magnetic field is initially weak during the linear stage, and the
kinetic spectrum has the Kolmogorov form. Magnetic fields grow
exponentially at the rate of the turbulent shear. Since the
viscous-scale eddies shear the fastest, magnetic growth proceeds at the
viscous timescale $t_\nu$.

The resistive scale of the magnetic field is determined by a balance
between shear growth and resistive decay: $t_s \sim \lambda_\eta^2 /
\eta$.  We identify the resistive scale $\lambda_\eta$ with the
transverse scale $\lambda_\perp$ (equation \ref{eq:sa}).  From the
beginning to the end of the nonlinear stage, $\lambda_\perp$ increases
by a factor of $(\lambda_f / \lambda_\nu)^{1/3}$, as is observed in
table \tablepower.

\subsection{Kulsrud-Anderson theory for the linear regime.}

Kulsrud and Anderson (1992) (also Kazantsev 1968, Gruzinov, Cowley,
Sudan 1996, Schekochihin, Boldyrev, Kulsrud 2001)
found that a dynamically weak magnetic field grows exponentially with
a $k^{3/2}$ spectrum terminating at the resistive scale.  The magnetic
spectrum evolves according to

\begin{equation}
\partial_t E_b(k)
  = \frac{\gamma}{5} \left( k^2 \partial_k^2 E_b(k) - 2 \partial_k
E_b(k) + 6 E_b(k) \right) - 2 \frac{k^2 \eta}{4\pi} E_b(k)
\end{equation}
which has the solution

\begin{equation} E_b(k,t) \sim k^{3/2} K_0 \left(\frac{5 \eta}{2\pi
\gamma} k\right) e^{(3/4)\gamma t} \end{equation}

where $\gamma \hh \sim \hh \int k E_v(k) d\!k$. $E_v(k) k^{3/2}$ is
the shearing rate at scale $k$.

We ran simulations starting from a weak magnetic field for a sequence
of 5 viscosities, all with Prandtl number $\ge 1$. Their ID numbers
are A1 through A5 (table \tablesima), and their energy and spectral
evolutions are shown in figures \figenergy\, and \figthreehalves.
Magnetic energies grow exponentially in all cases until the nonlinear
stage is reached. The growth rates $t_g$ are given in table
\tablepower, where they are expressed in the form $E_b(t) \sim
e^{t/t_g}$. The $k^{3/2}$ exponent is seen only when $P_r \ge 2500$,
as with simulation A1, where sufficient range exists between the
viscous and resistive scales (figure \figthreehalves). The $k^{3/2}$
spectrum is especially robust in the simulation with zero resistivity
(simulation A0, figure \figthreehalves). This simulation is not
without concern for its physical validity, a matter which we address
in sections \ref{foldscale}\,, \ref{dissipativepower}\,, and
\ref{refinement}.


\begin{center} {\it Table \tablepower: Energies and timescales}
\end{center}
\begin{center}
\begin{tabular}{ccccccccccccc}
\hline
Sim&$\nu$&$\eta$&$E_v$&$E_v$&$E_b$&$t_g$&$t_w$&$t_w$&$t_\eta$&$t_\eta$&
$s_\perp$&$s_\perp$\\
&&&\mbox{\tiny linear}&\mbox{\tiny sat}&\mbox{\tiny sat}&\mbox{\tiny linear}&
\mbox{\tiny linear}&
\mbox{\tiny sat}&\mbox{\tiny linear}&\mbox{\tiny sat}&\mbox{\tiny linear}&
\mbox{\tiny sat}\\
\hline
A1 &$5\!\cdot\!10^{-2}$&$2\!\cdot\!10^{-5}$&0.2 &.16&.17&1.41&1.5&1.55&
                       4.38&8.8&17.0&12.0\\
A2 &$5\!\cdot\!10^{-3}$&$1\!\cdot\!10^{-4}$&0.75&.38&.32&.59&.45&.62 &
                       1.50&3.5&13.0&8.5\\
A3 &$3\!\cdot\!10^{-3}$&$1\!\cdot\!10^{-4}$&0.8 &.40&.32&.46&.35&.47 &
                       1.26&3.5&14.2&8.5\\
A4 &$1\!\cdot\!10^{-3}$&$1\!\cdot\!10^{-4}$&0.9 &.60&.22&.36&.25&.30 &
                        .78&3.1&18.0&9.0\\
A5 &$4\!\cdot\!10^{-4}$&$4\!\cdot\!10^{-4}$&1.5 &.60&.16&.65&.12&.23 &
                        .30&1.8&14.6&6.0\\
\hline
B1 &$5\!\cdot\!10^{-2}$&$1\!\cdot\!10^{-5}$&   -&.16&.17&  -&  -&2.0&
                          -&12.9&   -&14.0\\
B2 &$5\!\cdot\!10^{-3}$&$4\!\cdot\!10^{-5}$&   -& .3& .4&  -&  -&.69 &
                          -& 4.4&   -&12.0\\
B3 &$3\!\cdot\!10^{-3}$&$4\!\cdot\!10^{-5}$&   -& .3& .3&  -&  -&.57 &
                          -& 4.4&   -&12.0\\
B4 &$1\!\cdot\!10^{-3}$&$4\!\cdot\!10^{-5}$&   -& .5& .3&  -&  -&.31 &
                          -& 3.5&   -&13.4\\
B5 &$4\!\cdot\!10^{-4}$&$1\!\cdot\!10^{-4}$&   -&.55& .3&  -&  -&.20 &
                          -& 3.4&   -&8.6\\
B6 &$1\!\cdot\!10^{-4}$&$1\!\cdot\!10^{-4}$&   -&.55& .3&  -&  -&.11 &
                          -& 3.1&   -&9.0\\
\hline
\end{tabular}
\end{center}

{\it \noindent ``Linear" and ``sat" denote the linear and satuated
states, averaged over a suitable length of time.  $E_v$ and $E_b$ are
the kinetic and magnetic energies. $t_g$ is the exponential magnetic
growth time during the linear phase.  $t_w$ is the vorticity time,
$t_\eta$ is the resistive time, and $s_\perp$ is the magnetic
wavenumber.  The $256^3$ simulations were run for half of a crossing
time, long enough to establish the saturated spectra and
$s_\parallel$, but not long enough to accurately determine $E_v$,
$E_b$, and $t_w$.}

The magnetic kurtosis, $<b^4>/<b^2>^2$, is observed to rise from $\sim
3$ to $\sim 15$ during the linear regime, and then to return to
$\sim 3$ in the saturated state. These results are discussed in
Schekochihin, Cowley, Maron, \& Malyshkin (2001).

For each value of the viscosity, the resistivity is assigned the smallest
value such that magnetic energy is destroyed by resistivity rather than
dealiasing.


\hbox to \hsize{ \hfill \epsfxsize7cm \epsffile{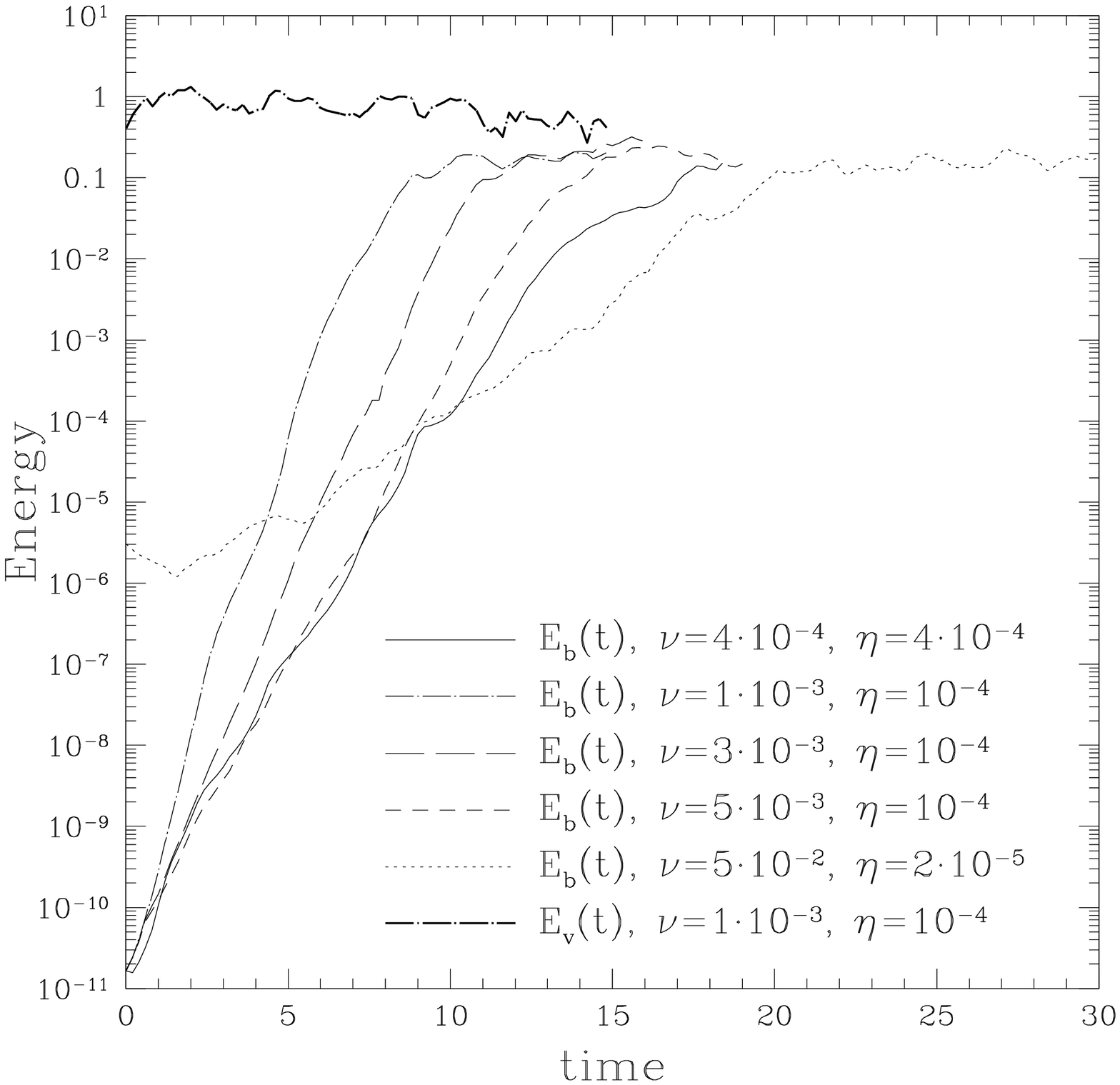} \hfill
\epsfxsize7cm \epsffile{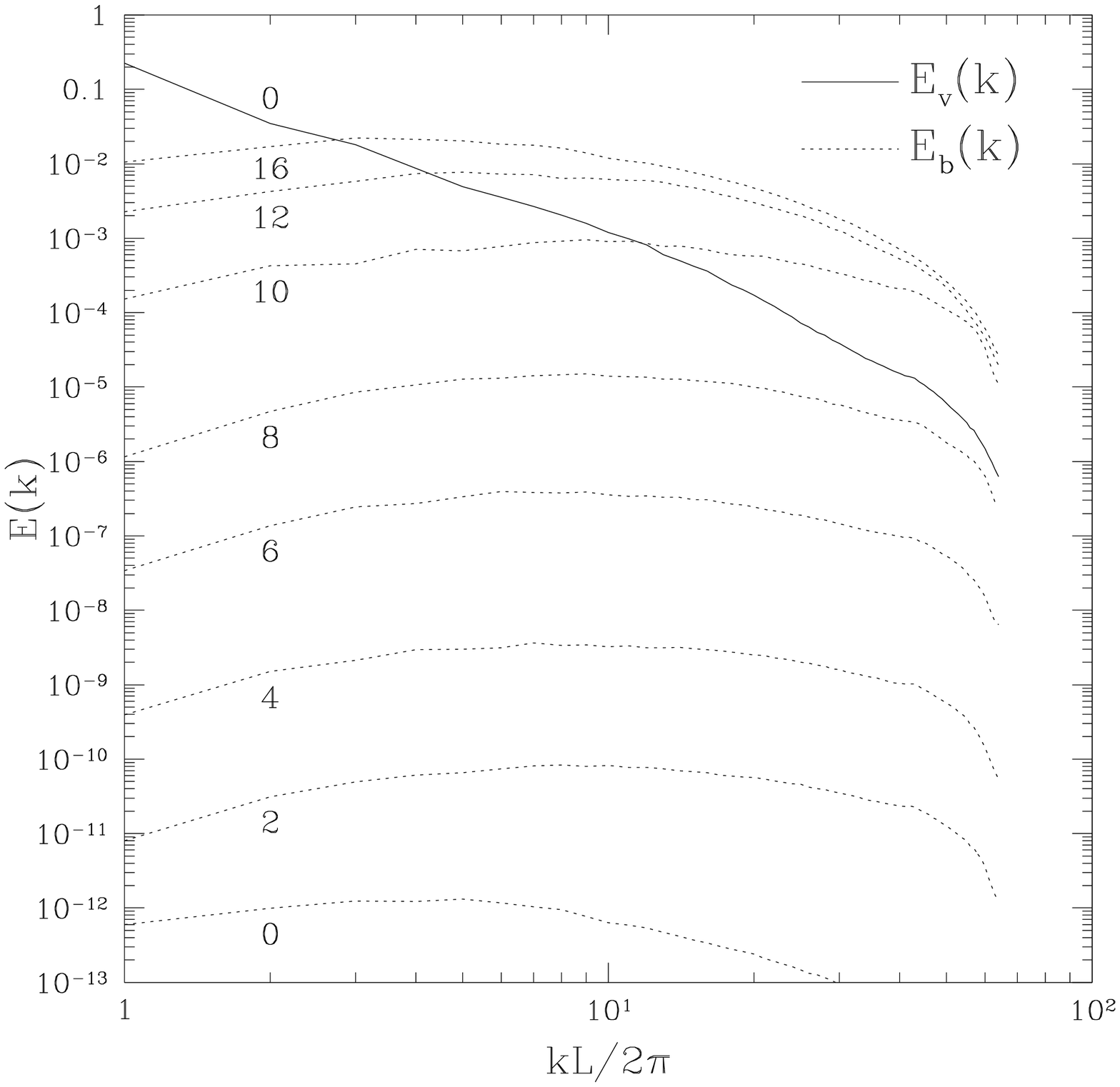} \hfill} \vspace{-4mm} \noindent {\it
Figure \figenergy. Left: The exponential magnetic energy growth
for a sequence of viscosities. Parameters for the simulation (A1
through A5) are given in table \tablesima. Right: The evolution of the
magnetic spectrum for simulation A3. Numbers indicate times. The
spectrum at the latest time is in the saturated state.}

\hbox to \hsize{ \hfill \epsfxsize7cm \epsffile{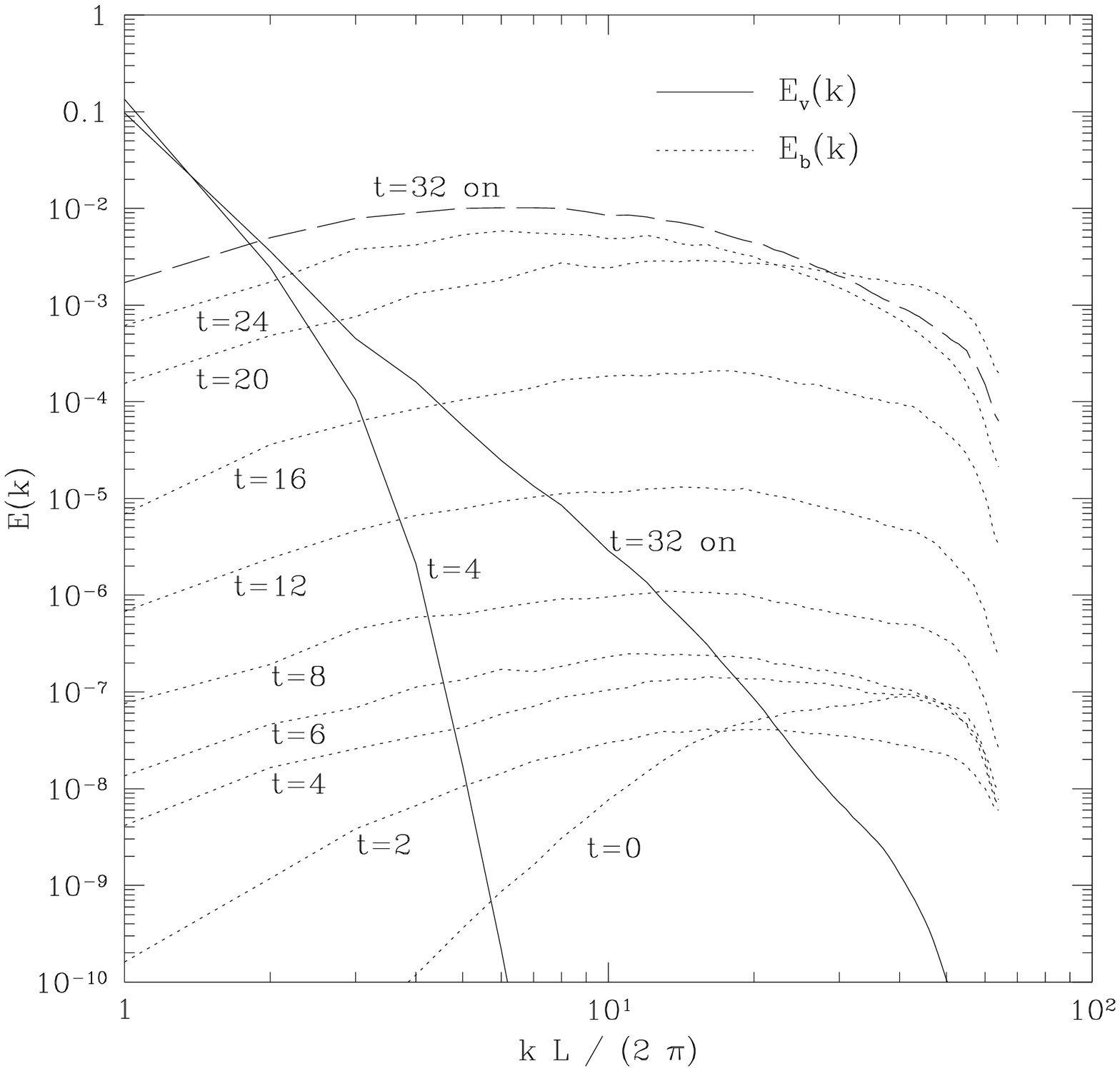} \hfill
\epsfxsize7cm \epsffile{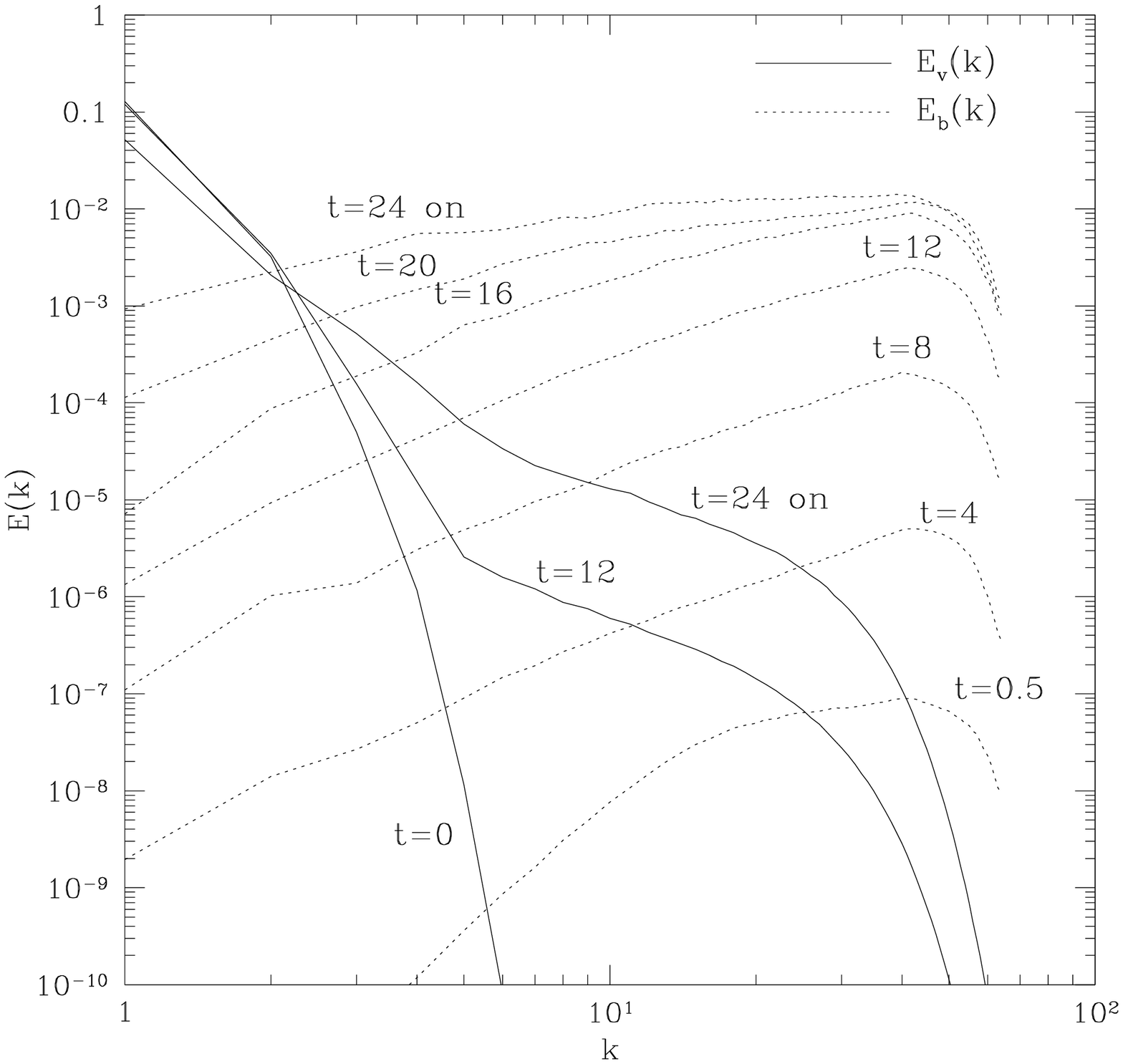} \hfill} \vspace{-6mm} \noindent {\it
Figure \figthreehalves.  Left: The kinetic and magnetic energy spectra for
simulation A1, $P_r = 2500$, $\nu=5\cdot10^{-2}$,
$\eta=2\cdot10^{-5}$. Right: The same quantities for simulation A0,
$\nu=5\cdot10^{-2}$, $\eta=0$. The Prandtl number is undefined,
however for this resolution and viscosity, any value above $10^{4}$ is
functionally equivalent.}

\section{Magnetic saturation} \label{Saturation}

Saturation is the long-term asymptotic state of MHD turbulence in the
absence of a mean magnetic field and with zero mean kinetic helicity.
We postulate that the dynamics are characterized by four principles:
1) The kinetic and magnetic energies are approximately equal. 2)
Magnetic energy forward cascades according to the forcing
timescale. 3) Magnetic fieldline folding and tangling prevents the
field from unwinding.  4) The magnetic field preferentially aligns
with the neutral shear frame.  These principles lead to a magnetic
spectrum of the form $k^{-1}$ that is independent of viscosity. The
spectrum is terminated at a resistive scale that is set by a balance
between forcing and resistivity: $\lambda_\eta \sim (t_f \eta)^{1/2}$.

The fastest kinematic shear is due to the viscous-scale eddies if the
kinetic spectrum is more shallow than $k^{-3}$, otherwise it is due to
the forcing-scale eddies. In the linear regime, where the kinetic
spectrum has the form $k^{5/3}$, the viscous-scale eddies shear
fastest. In this regime, shear functions to transfer energy from the
velocity to the magnetic field, and so the magnetic field grows at the
rate of the viscous-scale eddies.

In the nonlinear regime, the magnetic backreaction opposes shear that
is less energetic than the magnetic field. This effect can be nonlocal
in Fourier space, where small-scale magnetic fields can oppose large
scale shear. In the saturated state, where the kinetic and magnetic
energies are equal, only the forcing-scale eddies are energetic enough
to shear the magnetic field.  Shear exists on smaller time and energy
scales, yet we argue that some fraction of it results from a transfer
of magnetic energy to the velocity, rather than the other way around.
Or, energy can be cyclically exchanged between $\vv$ and $\bb$, such
as for spandex waves (section \ref{spandex}).

The magnetic field has a reduced tension per energy compared to what
would be expected from random phase structure. The reduction is
typically by a factor of $10$ for a $128^3$ simulation, as evaluated
by the parameterization in section \ref{structure}. Therefore, less
generation of small scale velocities by the unwinding of small scale
magnetic fields occurs. The reduction is great enough so that the
magnetic energy cascades from the forcing scale to the resistive scale
without significant loss to unwinding. This phenomenon is exhibited in
the tension release simulations of section \ref{unwind}, and is
reflected by the independence of the magnetic spectrum on viscosity.

In the simulations with an initially weak magnetic field, the magnetic
spectrum is dominated by small scale energy at the beginning of the
nonlinear stage, and in subsequent evolution it stays this way. One
may ask if the result would be different if we started instead with a
large scale, large energy field. Any small scale field that is formed
should because the fieldline structure is not yet folded and the
tension is strong. We observe instead that the magnetic struture
evolves to a small-scale dominated state identical to the result of a
weak initial field simulation. The likely reason for this is that
fieldlines tangle, and even if they are energetically able to unwind,
topological constraints prevent them. In general, we have never
observed magnetic hysteresis in any of the simulations.

Figure \figfinalstate\, illustrates the nonlinearly saturated spectra
for a sequence of viscosities, all at fixed resistivity. The magnetic
spectra all fall in the same place, with the exception of the one
having the highest viscosity. This suggests that the kinematics at any
scale other than the forcing scale are irrelevant to the cascade. The
kinetic spectra also have greatly reduced magnitudes compared to the
magnetic spectra at high $k$.

\hbox to \hsize{ \hfill \epsfxsize7cm \epsffile{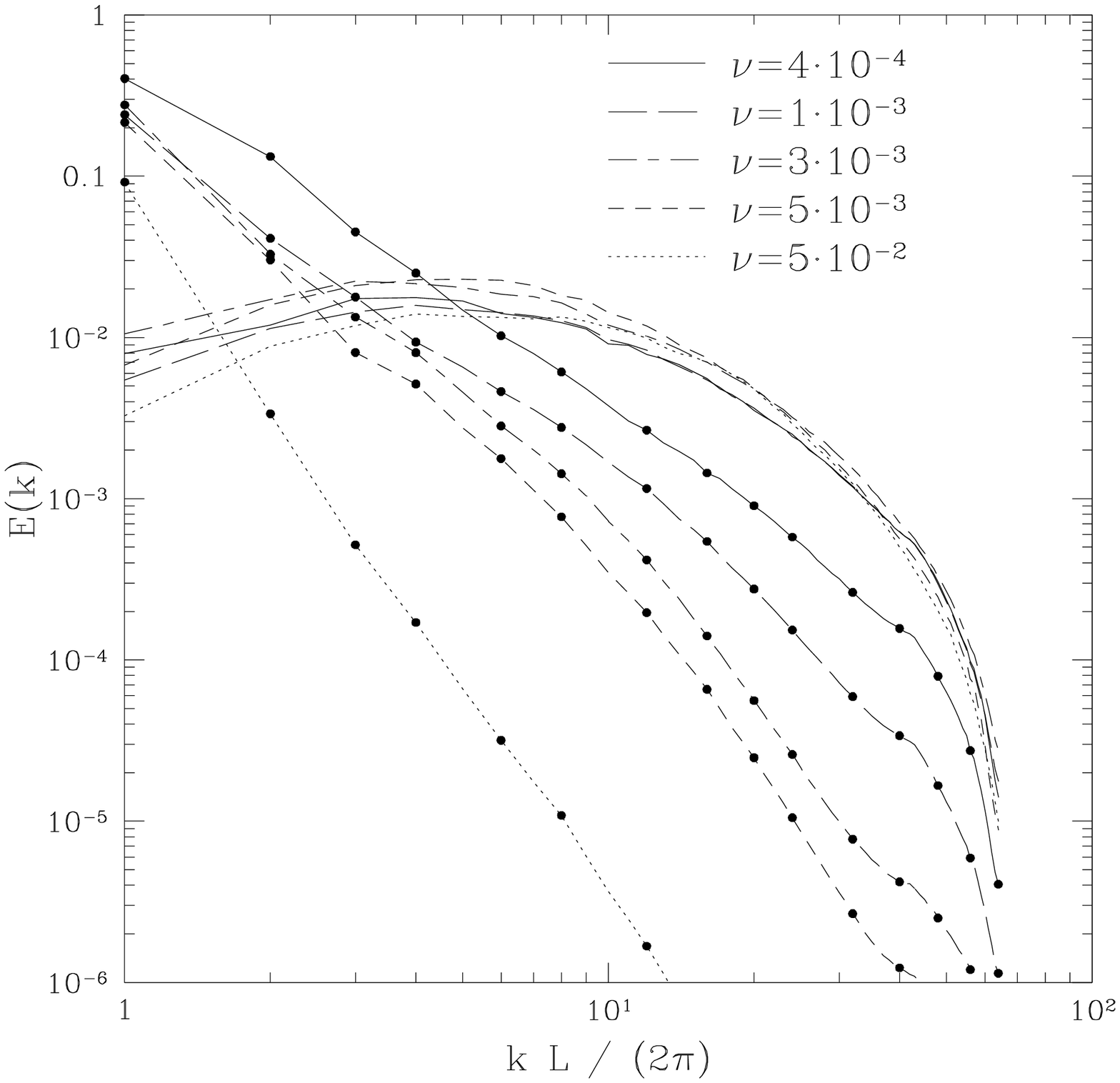}
\hfill \epsfxsize7cm \epsffile{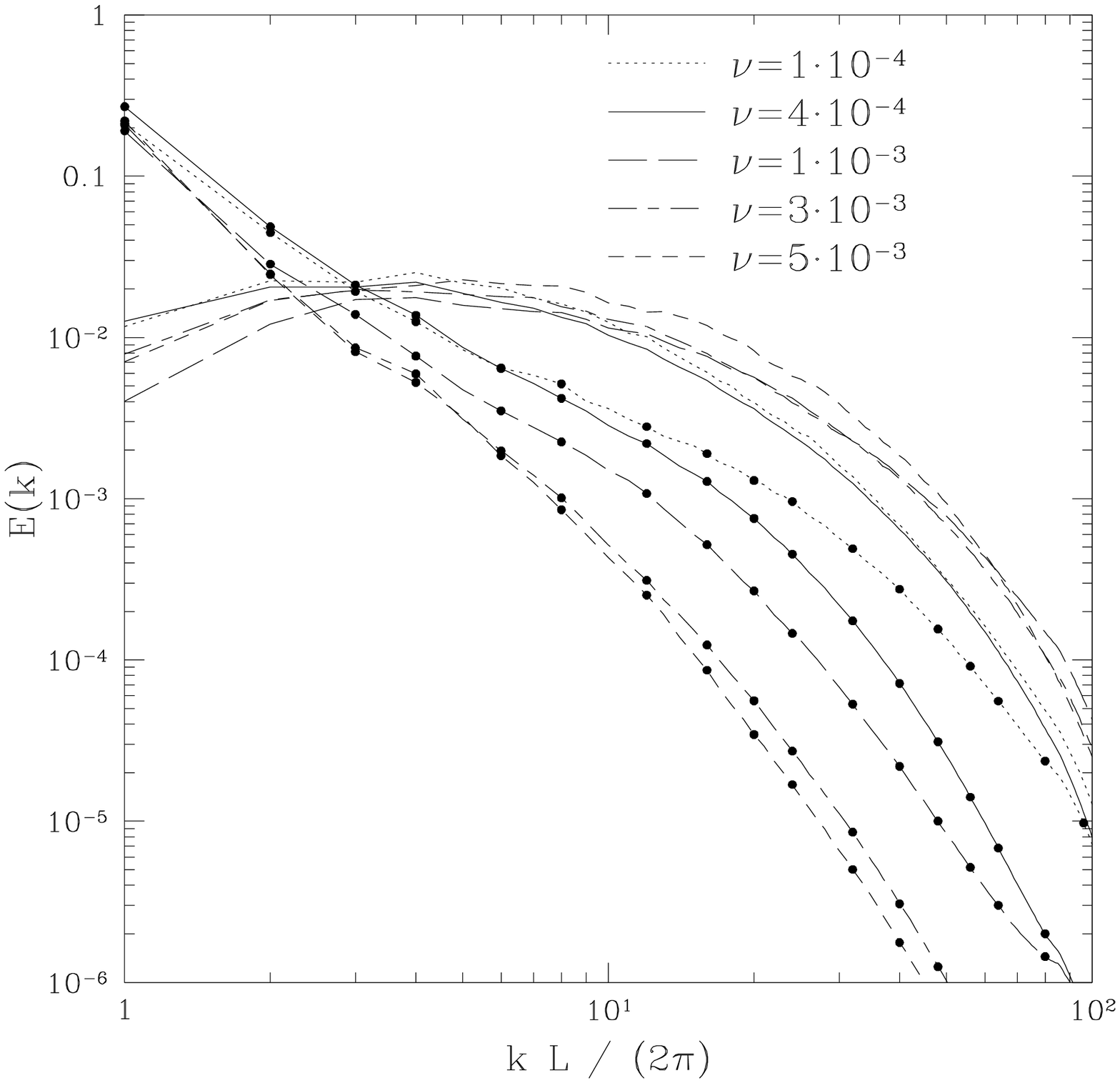} \hfill}
\vspace{-4mm} \noindent {\it Figure \figfinalstate: The saturated
spectra for a sequence of viscosities and fixed resistivity. Lines
with points indicate the kinetic spectra, and lines without points
indicate the magnetic spectra. The simulations in the left and right
panels have $128^3$ and $256^3$ resolution, respectively.  The ID
numbers for the $128^3$ simulations are A1 through A5, and for the
$256^3$ simulations they are B2 through B6.}

\subsection{Magnetic growth during the nonlinear phase}

During the linear stage, the magnetic energy is less than the
viscous-scale energy $(b < v_\nu),$ and the velocity spectrum has the
Kolmogorov form. The fastest shear ($t_s$) is due to the viscous-scale
eddies: $\lambda_s \sim \lambda_\nu,$ $v_s \sim v_\nu,$ and $t_s \sim
t_\nu.$ The nonlinear stage begins when $b \sim v_\nu$. Thereafter,
growth slows as the shear velocity grows according to $v_s \sim b.$ At
and above the shear scale, the velocity still has the Kolmogorov form,
and the shear time is $t_s \sim \lambda_s / v_s \sim \lambda_f v_s^2 /
v_f^3 \sim\lambda_f b^2 / v_f^3$. Growth ends when $b \sim v_f$ and
$t_s \sim t_f$.

The resistive scale of the magnetic field is determined by a balance
between shear growth and resistive decay: $t_s \sim \lambda_\eta^2 /
\eta$.  We define $\lambda_\eta = \lambda_\perp$ (equation
\ref{eq:sa}).  From the beginning to the end of the nonlinear stage,
$\lambda_\perp$ increases by a factor of $(\lambda_f /
\lambda_\nu)^{1/3}$. The data in table \tablepower supports this
scaling, although comparison is inexact because the forcing velocity,
and hence the shear time, change slightly from the linear to the
saturated state.

The magnetic enegy grows during the nonlinear stage in our 3
dimensional simulations, whereas in the 2 dimensional simulations of
Kinney et. al. 2000, it decays. In 2D, they found that an initially
weak field grows exponentially at the resistive scale until
nonlinearity occurs, after which the field decays.

\subsection{Timescales and the neutralization of shear}

According to the backreaction hypothesis, shear motions below the
forcing scale have a dimished role in cascading the magnetic field. If
true, then the vorticity time $t_w$ reflects some shear motions not
associated with the cascade of the magnetic field, whereas the
resistive time $t_\eta$ more accurately reflects the magnetic cascade
time.  We therefore expect a greater change in $t_\eta$ from the
linear to the saturated state than for $t_w$, and we also expect
$t_\eta$ to be more approximately constant as a function of viscosity
than $t_w$. Both of these conclusions are well supported by the data
in table \tablepower.  $t_\eta$ is not expected to be exactly constant
in the saturated state because the larger viscosities influence the
forcing-scale velocities, and hence $E_v$ and the shear rate.

\subsection{Large initial magnetic field} \label{largek}

Kulsrud-Anderson predict that an initially weak field grows
exponentially with a $k^{3/2}$ spectrum, which simulation
confirms. The $k^{3/2}$ spectrum is independent of the initial
spectral shape.  The magnetic energy is therefore dominated by
small-scale structure at the beginning of the nonlinear phase, and
remains so through subsequent evolution to the saturated state.  One
may ask if the result would be different if the magnetic field were
initially strong and organized at low $k$. We will find that there is
no hysteresis in the saturated state from the results in this section,
and in section \ref{refinement}\, on resolution refinement.

Simulation L3 has an initial magnetic energy of unity, which is
confined to modes with $k/2\pi \leq 4$. The viscosity is
$3\cdot10^{-3}$, and the resistivity is $10^{-4}$, both of which are
identical to simulation A3. The viscous scale is smaller than the
initial magnetic scale. Subsequent evolution restores the magnetic
spectrum to the saturated state of simulation A3, as show in figure
\figbigb.

A more stringent test is to initialize a strong large-scale magnetic
field which has no topological linkage. Such a field can resist
bending by the turbulence and possibly oppose the formation of small
scale field.  If the field is topologically linked, then even though
tension forces exist, they may be unable to release the tension.  The
question is, do linkages form in a strong magnetic field?

The field of $b_2 = 2 \sin (2\pi x_1)$ is a simple example having zero
mean and an energy density of unity, and constitutes the initial
conditions of simulation L4. The initial velocity is zero, and the
forcing power is unity. The viscosity is $10^{-3}$ and the resistivity
is $10^{-4},$ which are identical to simulation A4. The magnetic
field evolves to the saturated state A4 after 5 time units.

\hbox to \hsize{ \hfill \epsfxsize7cm \epsffile{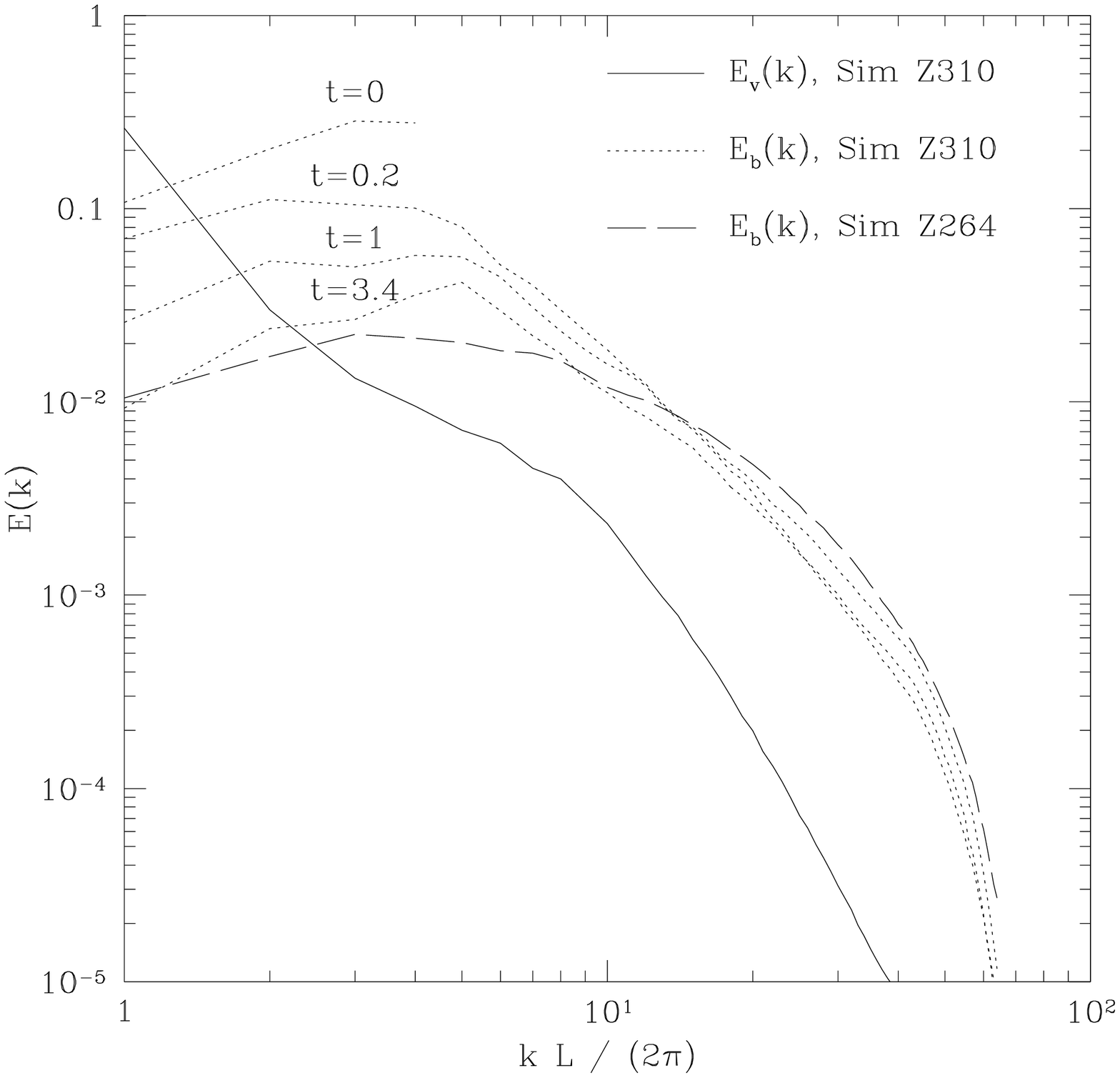} \hfill
\epsfxsize7cm \epsffile{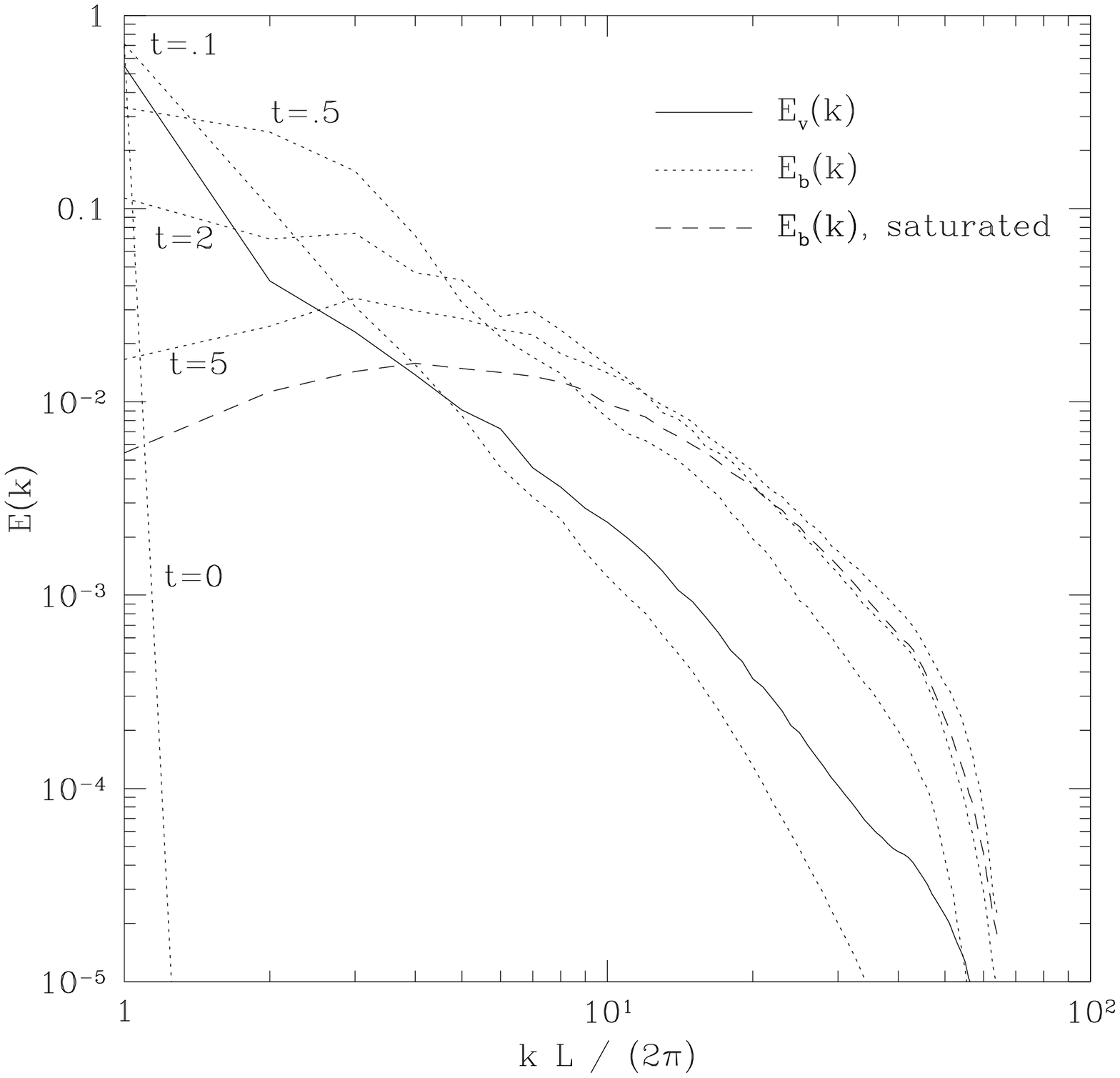} \hfill} \vspace{-4mm} \noindent
{\it Figure \figbigb. Left: this figure follows the evolution of an
initially large magnetic field at low $k$ (simulation L3), until it
reaches a state where the magnetic energy is dominantly at high
$k$. The dashed line is the time-averaged saturated state of
simulation A3, which has the same viscosity and resistivity.  Right:
This figure follows simulation L4, which is similar to L3 except
that the initial field is at even lower $k$.  The dashed line is the
time-averaged saturated state of simulation A4, which has the same
viscosity and restivity.}

\subsection{Interaction between the velocity and the magnetic field}
\label{unwind}

The reduced magnetic tension of the saturated state was previously
diagnosed by noting that $\lambda_\parallel/\lambda_\perp >> 1$. A
more direct test can be constructed by observing how fast magnetic
tension generates kinetic energy. This is done by artificially setting
the velocity to zero and observing the release of tension in the
subsequent evolution. Specifically, we note how much new kinetic
energy is generated from fieldline unwinding, and where in Fourier
space it appears.  Folded fieldlines have less tension per energy than
structureless fieldlines, and hence unwind more slowly, generating
less velocity.  The random-phase magnetic field defined in appendix
\ref{randomphase} serves as a reference for structureless
fieldlines. A random-phase field is generated from a structured field
by randomizing the Fourier component orientations while preserving the
power spectrum. The structured magnetic field generates more velocity
than the random-phase field, verifying that it has reduced tension per
energy (figure \figrelease).

We draw from the nonlinearly saturated state of simulation A4 at
$t=12$ to initialize two test simulations.  In simulation U4, we
erased the velocity and restarted with the same viscosity. The
resistivity was reduced to $10^{-5}$ to remove the effect of resistive
magnetic energy loss. In simulation U4r we additionally randomize the
phases of the magnetic Fourier modes. A random phase magnetic field
has a folding factor of unity, and serves as a reference for assessing
structure.  By comparing the kinetic spectra generated in simulations
U4 and U4r, we can evaluate the degree of folding present in U4.

We also observe that small scale magnetic fields generates large scale
velocities, while the random-phase field generates small scale
velocities. Specifically, the original magnetic field at $s \sim 15$
generates velocities at $s \sim 3$, while the random phase field
generates them at $s \sim 15$. We infer that for the dynamics of the
saturated state, only large scale shear is responsible for the
magnetic cascade, even though the kinetic spectrum is more shallow
than $k^{-3}$. Therefore, the cascade proceeds according to one
timescale, the forcing timescale. This also supports our claim that
magnetic forces oppose shear that is less energetic than the magnetic
field. Finally, the absence of production of kinetic energy beyond
$s=6$ establishes that unwinding of the small scale field is
unimportant. The magnetic energy cascades from the forcing scale to
the viscous scale without loss to the velocity.

\hbox to \hsize{ \epsfxsize7cm \epsffile{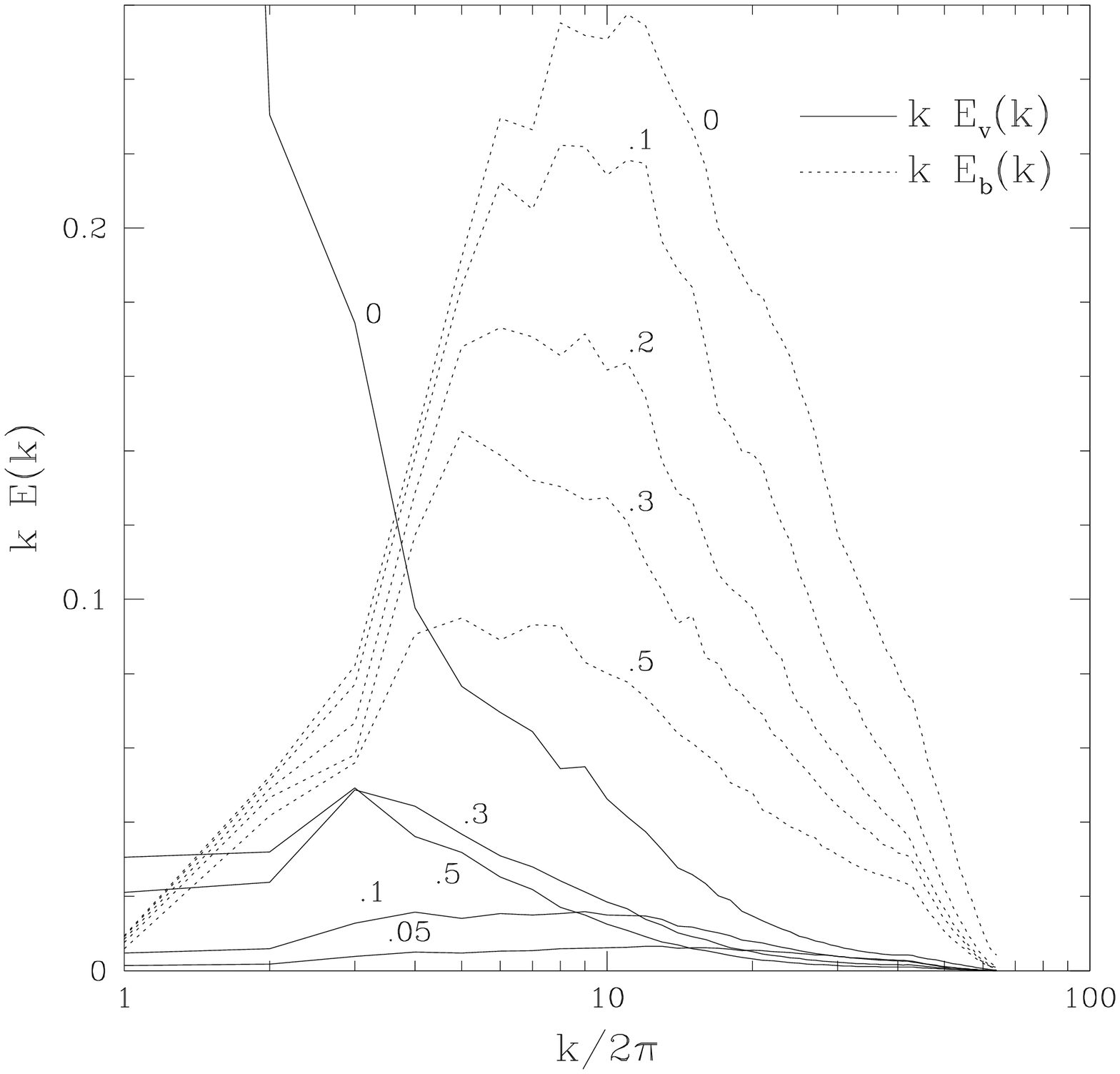} \hfill
\epsfxsize7cm \epsffile{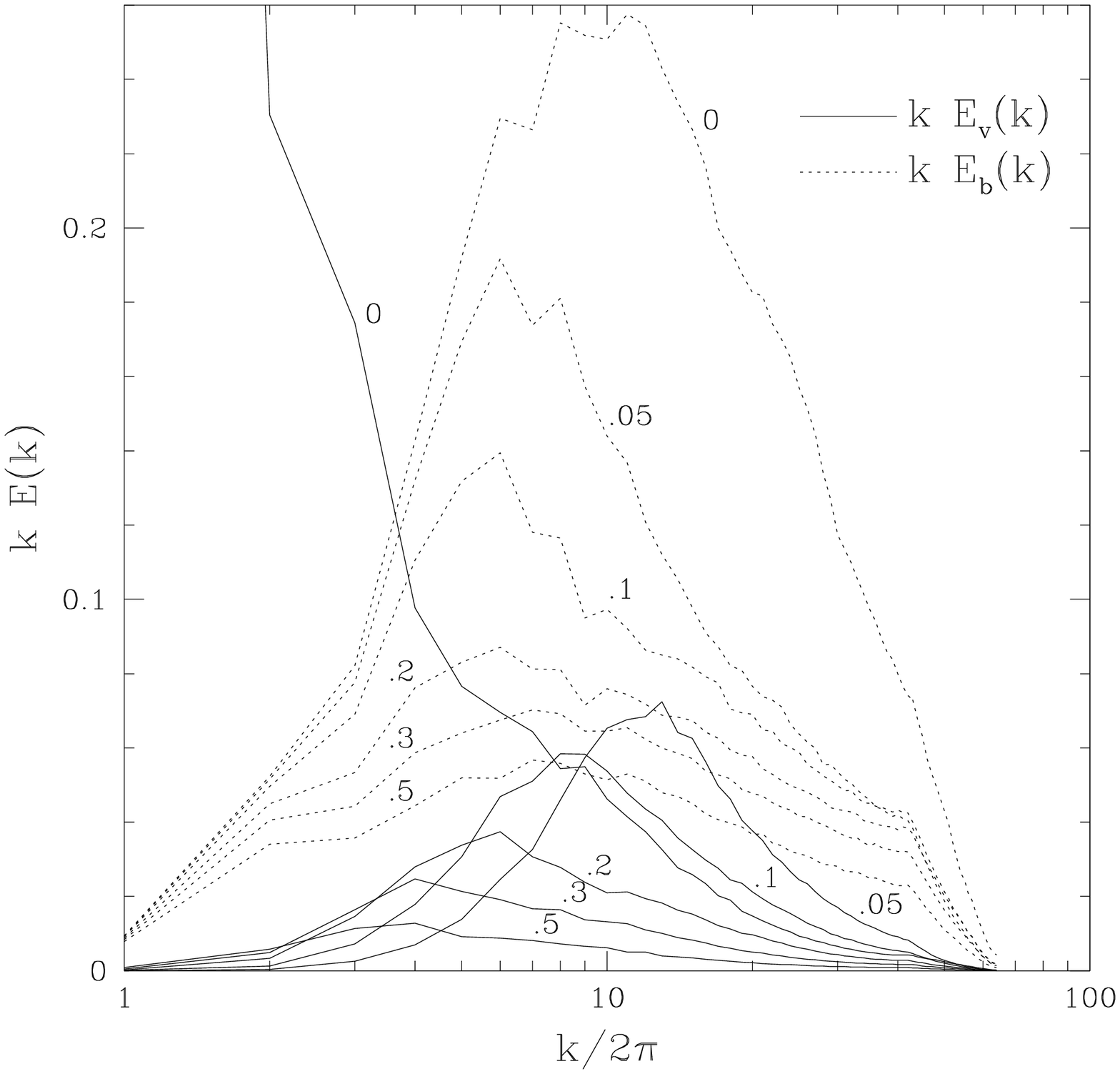} } \vspace{-4mm}
\noindent {\it Figure \figrelease. Left: Simulation U4. Right:
Simulation U4r. Numbers indicate times.  These simulations are
discussed in section \ref{unwind}. At $t=0$, we show the kinetic
spectra from before the velocity was artificially set to zero.}

\subsection{Folding scales} \label{foldscale}

We evaluated the longitudinal and transverse folding scales
$\lambda_\parallel$ and $\lambda_\perp$ of the saturated states for a
range $\nu$ and $\eta$ (figure \figfoldscale).  For a $128^3$
simulation, and for $\eta$ larger than $10^{-4}$, we find that
$\lambda_\perp \sim \lambda_\eta \sim \eta^{1/2}$, which would be
expected if the shear timescale is constant and shear balances
resistivity. $\lambda_\parallel$ is approximately constant and equal
to the outer scale of unity. For $\eta < 10^{-4}$, decreasing the
resistivity does not further decrease $\lambda_\perp$, but it does
decrease in $\lambda_\parallel$ and the folding factor. This occurs
because for these low values of $\eta$, dealiasing of the magnetic
field disrupts folded structure (section \ref{refinement}). We
interpret the lowest attainable value of $\lambda_\perp$ as the
resolution limit.

For a $256^3$ simulation, the resolution limit is reached at $\eta
\sim 4\cdot 10^{-5}.$ At this resolution, $\lambda_\perp$ is smaller
than for a $128^3$ simulation, while $\lambda_\parallel$ is unchanged.
Further reduction of $\eta$ does not further reduce $\lambda_\perp.$

The saturated spectra for varying resistivity at fixed viscosity are
generated by a process of resaturation, whereby we take a saturated
state, change the resistivity, and continue evolution until the system
has reached a new equilibrium state. This process is described in
section \ref{refinement}. The initial conditions are always taken from
one of the simulations in the series A1 through A5. $256^3$
simulations are initialized by doubling the grid of the appropriate
$128^3$ simulation, resetting the resistivity, and then evolving to
the new saturated state.

\hbox to \hsize{ \hfill \epsfxsize7cm \epsffile{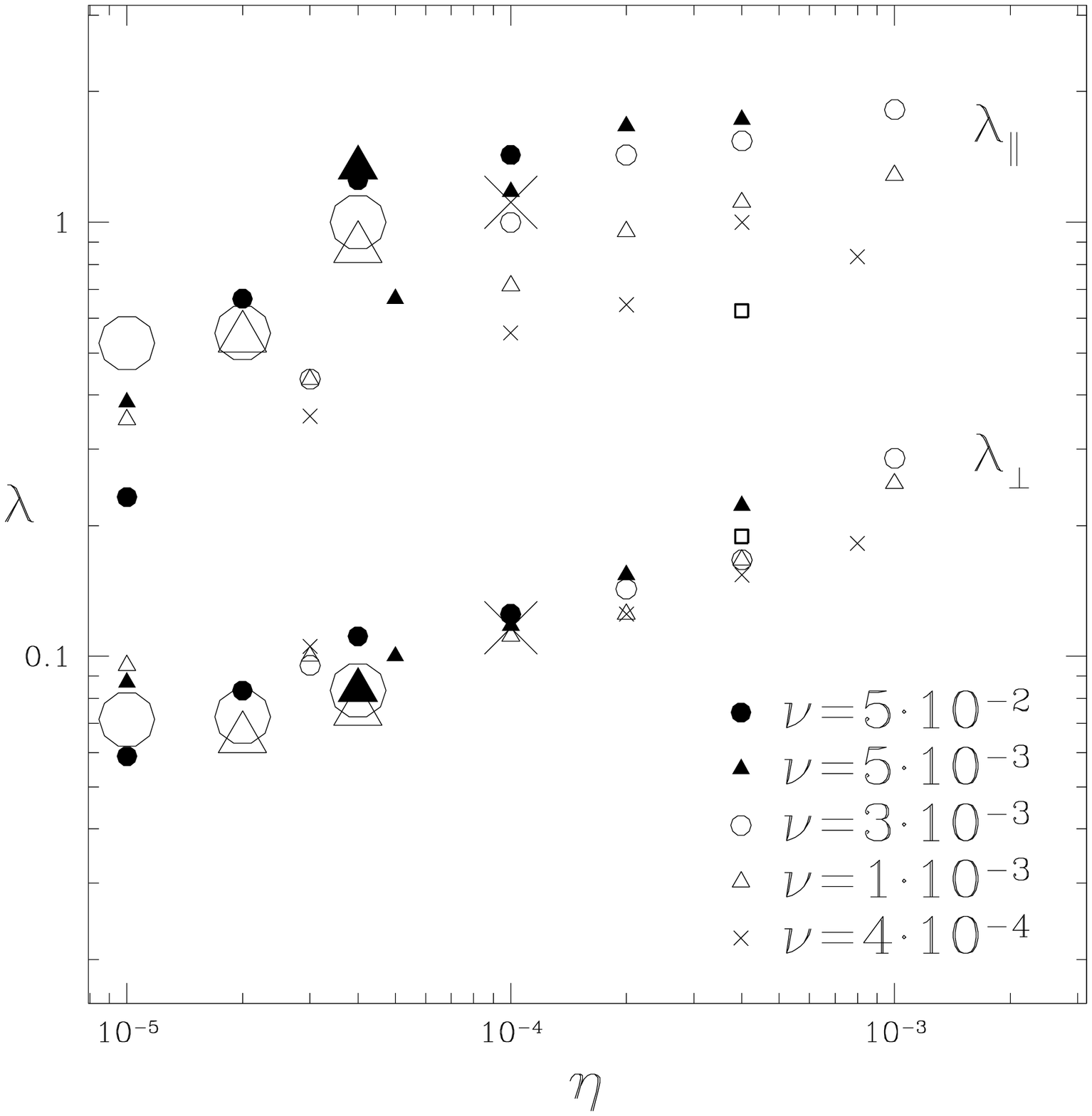}
\hfill} \vspace{-4mm} \noindent {\it Figure \figfoldscale: The upper
and lower rows of points correspond to the longitudinal and transverse
folding scales defined in section \ref{structure}. These numbers come
from the saturated states of simulations for a range of values in
$\nu$ and $\eta$. Small symbols represent $128^3$ simulations, and
large symbols represent $256^3$ simulations. The square points in the
figure represent a $64^3$ simulation with a Braginskii viscosity of
$\nu_T=.003$, and with $\eta=.0004$. This establishes that fieldline
structure is folded in this simulation.}

\subsection{Unwinding}

For scales large enough to be uninfluenced by viscosity, magnetic
fieldlines release tension in one Alfv\'en time: $t_A \sim
\lambda_\parallel / b.$ Below the viscous scale, the viscous and the
magnetic tension terms dominate in the Navier Stokes equation: $\nu
v_u / \lambda_\perp^2 \sim b^2 / \lambda_\parallel$, where $v_u$ is
the velocity at which magnetic structures zunwind.  The
viscous unwinding time is then $t_u \sim \lambda_\parallel / v_u \sim
\lambda_\parallel^2 \nu / (\lambda_\perp^2 b^2)$.  We have accounted
for the possibility of folded magnetic structure by distinguishing
between $\lambda_\perp$ and $\lambda_\parallel$.

Magnetic energy is lost Alfv\'enically if $t_A > t_u$, otherwise it is
lost viscously or resistively, whichever is faster. The transition
between Alfv\'en and viscous unwinding occurs at a scale $\lambda_A$
such that $t_A \sim t_u.$ Assume a spectrum of $E_b(k) \sim k^{-1}$,
or $b \sim \lambda^0.$ We consider two limiting cases for the folding
structure.  If $\lambda_\parallel \sim \lambda_f$, then $\lambda_A
\sim \lambda_f^{1/3} \lambda_\nu^{2/3}$. If $\lambda_\parallel \sim
\lambda_\perp$, then $\lambda_A \sim \lambda_\nu^{4/3}
\lambda_\f^{-1/3}.$

At small scales, where viscosity mediates unwinding, we ask which is
faster, shear or unwinding.  If $t_s < t_u$, then the magnetic energy
cascades without loss to the resistive scale, which is determined by a
balance between shear and resistivity: $\lambda_\eta \sim (t_s
\eta)^{1/2}.$ The magnetic energy would then grow to be equal to the
forcing energy.  If $t_s > t_u$, then the magnetic energy must adjust
itself to restore $t_s \sim t_u \sim t_\eta$.  We define the resistive
scale to be equal to the transverse magnetic scale: $\lambda_\eta =
\lambda_\perp$.

The ratio of the unwinding to the shear time is \begin{equation}
\frac{t_u}{t_s} \sim \frac{\lambda_\parallel^2}{\lambda_s^2}
\frac{v_s^2}{b^2} \frac{\nu}{\eta} \label{abc} \end{equation}
Simulation suggests that the magnetic field and the shear eddies have
the same energy: $b^2 \sim v_s^2$, and that the folding scale is such
that $\lambda_\parallel \sim \lambda_f$. We assume that the Kolmogorov
scaling applies above the shear scale: $v_s^3 \lambda_f \sim v_f^3
\lambda_s.$

If $t_u > t_s$, then $b \sim v_f$ and $\lambda_s=\lambda_f$, and
consistency with (\ref{abc}) requires that \begin{equation} \Lambda
\equiv \frac{\nu}{\eta} \frac{\lambda_\parallel^2}{\lambda_f^2} \geq
1.  \end{equation}

If $\Lambda < 1$, then unwinding balances shear, and $b^2$ decreases
to bring about $t_u \sim t_s \sim t_\eta$. The shear time and the
magnetic scale both contract until $\lambda_\eta \sim (\eta
t_f)^{1/2}\, \Lambda^{1/6}.$ The magnetic energy is then $b^2 \sim
v_f^2 \Lambda^{1/3}.$

\subsection{Resolution refinement} \label{refinement}

We found in section \ref{largek} that the saturated state is free from
hysteresis. There, we compared the final state of simulation A4, which
started from weak magnetic field conditions at high k, with that of
simulation L4, which started from strong magnetic field conditions at
low k. The viscosity and resistivity are the same in each of these
simulations. In this section, we consider the effect of changing the
resistivity.

In a process which we call resaturation, we take the saturated state
from a finished simulation, change the resistivity, and then continue
the simulation until a new saturated state is reached. Simulations A4s
and A5s were thus obtained from simulations A4 and A5, respectively.
In simulations A4s and A5s, the resistivity was reduced to $10^{-6}$.

Simulations A4w and A5w have the same $\nu$ and $\eta$ as A4s and A5s,
but they were started instead from a weak initial magnetic field with
an energy of $10^{-6}$. The saturated spectra for A4s and A4w are
identical, as are the ones for A5s and A5w.

The new saturated states for simulations A4s and A5s were reached in a
time equal to one third of the forcing time. Whereas the technique of
resaturation has now been validated, we can thus obtain the saturated
states for any combination of $\nu$ and $\eta$ with little
computational effort. This procedure was exploited to obtain the data
used in section \ref{foldscale}, and also to obtain the saturated
spectra of the $256^3$ simulations B1 through B6.

Simulation A4s develops only slightly more small-scale magnetic field
than simulation A4. This is because the resistivity is small enough so
that the spectral aliasing procedure contributes to magnetic energy
loss, which destroys folded fieldline structure and increases the
magnetic tension per energy. We show this with the tension spectrum in
figure \figfoldspectrum. The extra magnetic forces generate velocities
which dissipate viscously, removing magnetic energy.  Viscosity
takes over the role of resistivity, diminishing the effect of lowering
the resistivity in simulation A4s.

Although not obvious in the spectra, this effect can be seen by
plotting the viscous and resistive dissipation as in figure
\figbfcloss.  The aliasing scale is reduced by a factor of 2 in a
$256^3$ simulation, where it then does not trigger viscous
dissipation.

\hbox to \hsize{ \hfill \epsfxsize7cm \epsffile{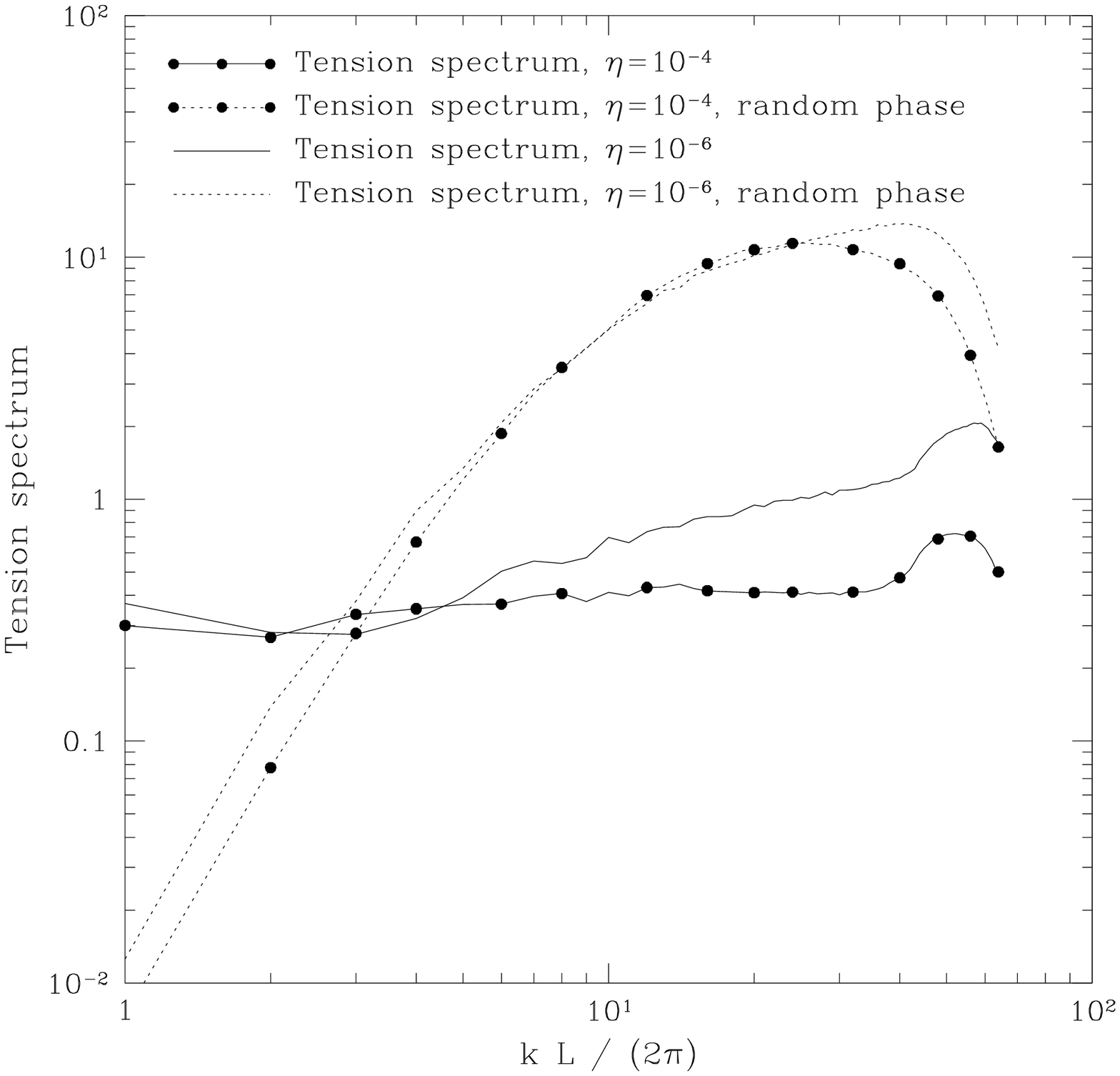}
\hfill} \vspace{-4mm} \noindent {\it Figure \figfoldspectrum: We plot
the spectrum of the tension term, $\bb\cdot\nabla \bb,$ to show the
effect of lowering the resistivity enough so that aliasing destroys
folding structure and enhances tension. The simulations are A4 and
A4w, which both have $\nu=10^{-3}.$ A4 has a resistivity of $10^{-4}$,
and A4w has a resistivity of $10^{-6}$. The random phase spectra are
obtained by the random phase transformation defined in section
\ref{randomphase}}

\subsection{Dissipative power} \label{dissipativepower}

We compare the spectral distribution of energy and dissipation for the
saturated state by plotting each quantity logarithmically distributed
in $k$. For the kinetic and magnetic spectra, we plot $\ln(10) k
E(k)$, and for the viscous and resistive power, $2 \nu \ln(10) k^3
E_v(k)$ and $2 \eta \ln(10) k^3 E_b(k).$ When plotted on a linear
ordinate, these curves have the interpretation that the total quantity
is equal to the area under the curve.


In figure \figbfcloss, it is clear for simulation A4 that the magnetic
energy resides small scales, and that the high-k cutoff is due to
resistivity. Lowering the resistivity by a factor of $3.3$ in
simulation A4t does not significantly change the magnetic energy
spectrum because viscosity takes the place of resistivity at high $k$.
The interpretation is that in simulation A4t, the resistivity is
sufficiently low so that aliasing contributes to magnetic energy loss.
This destroys magnetic fieldline folding at high $k$, increases the
tension (figure \figfoldspectrum), and leads to more viscous
dissipation. A proper study of the effect of lowering the resistivity
requires that we simultaneously increase the resolution.

\hbox to \hsize{ \hfill \epsfxsize7cm \epsffile{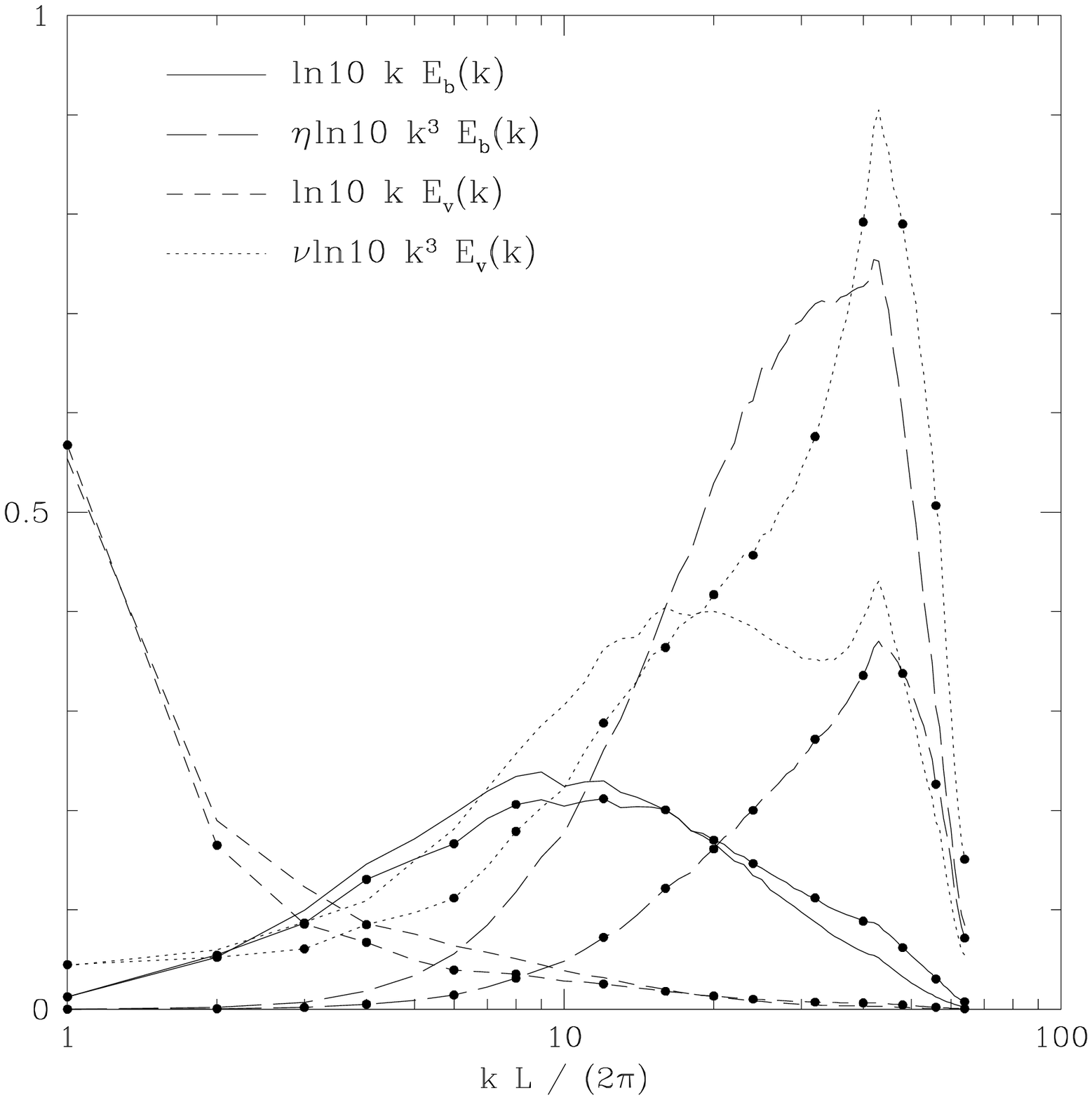} \hfill
\epsfxsize7cm \epsffile{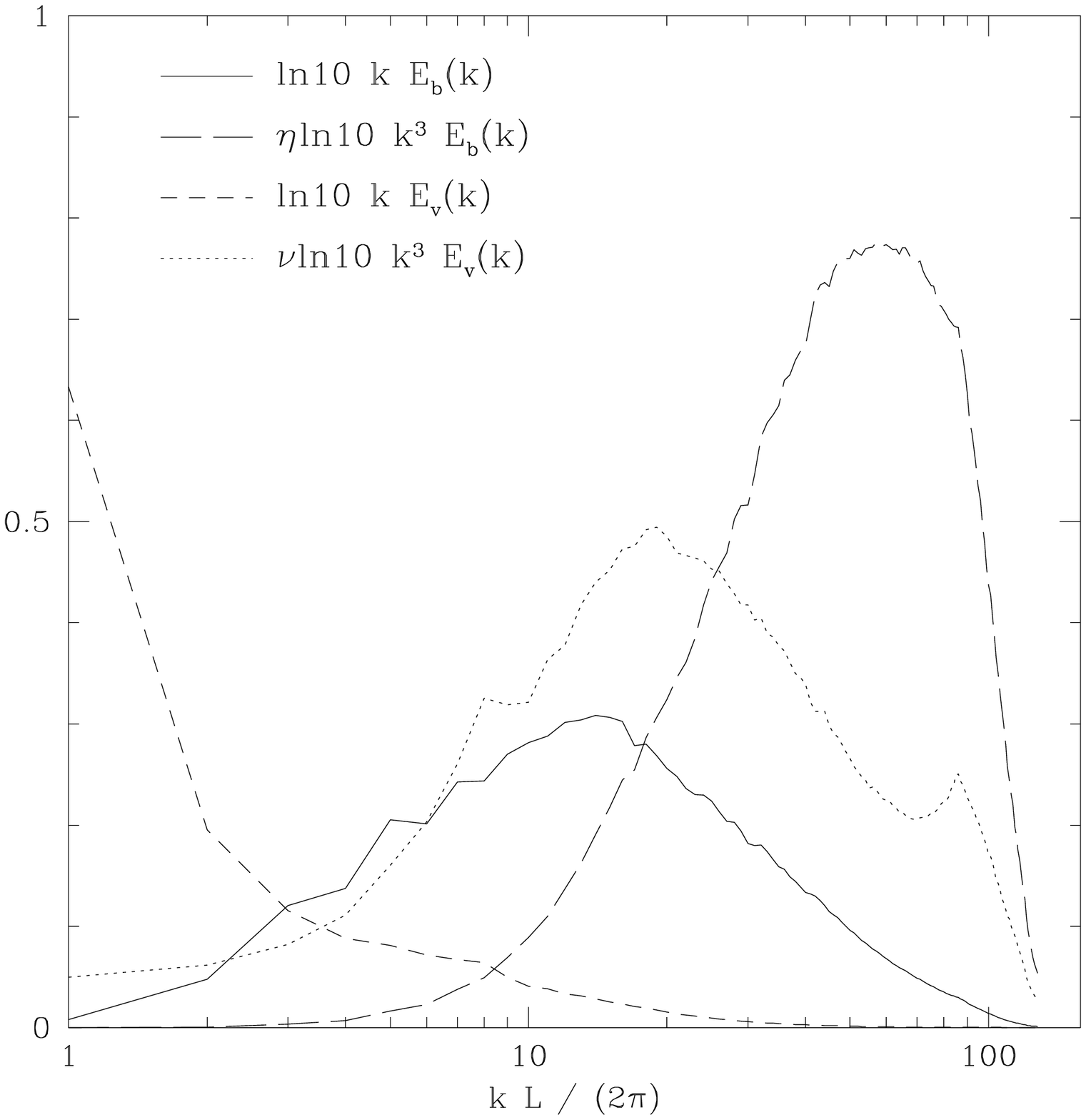} \hfill} \vspace{-6mm} \noindent
{\it Figure \figbfcloss. Left: We plot the energy and dissipation per
logarithmic $k$ for the velocity and magnetic fields. Lines without
dots denote simulation A4, with $\nu=10^{-3}$ and
$\eta=10^{-4}$. Lines with dots denote simulation A4t, which has the
same viscosity and has $\eta=3\cdot10^{-5}$. Right: The energy and
dissipation per logarithmic $k$ for the $256^3$ simulation B4, which
has $\nu=10^{-3}$ and $\eta=4\cdot10^{-5}$. The magnetic spectrum of
simulation B4 farther toward the right in k than for simulation A4.}

\subsection{Shear and magnetic alignment} \label{shear}

The symmetric shear matrix $S = (\partial_i v_j + \partial_j v_i) / 2$
can be diagonalized to yield the shear eigenvalues
$\{\lambda_+,\lambda_f,\lambda_-\}$ and eigenvectors
$\{s_+,s_0,s_-\}$, where we list the eigenvalues from largest to
smallest. Incompressibility ensures that $\lambda_+ + \lambda_f +
\lambda_- = 0.$ The magnetic energy growth is $G = .5 <\partial_t b^2>
= <b_i b_j S_{ij}>.$ With the magnetic field projected into the local
shear eigenvector frame, this is $G = <b_+^2\lambda_+ + b_0^2\lambda_f
+ b_-^2\lambda_->.$

In a nonmagnetized fluid, let a passive scalar $c$ have a diffusivity
$\zeta$ which is substantially less than $\nu$. $\nu/\zeta$ is known
as the Schmidt number.  In the range between $\lambda_\nu$ and
$\lambda_\zeta$, where $\lambda_\zeta$ is the passive scalar diffusion
scale, the passive scalar cascade rate is $\epsilon_c \, \sim \, c^2 /
(\lambda_\nu/v_\nu)$, leading to a steady-state cascade index of
$E_c(k) \, \sim \, k^{-1}$.

Consider two extremes for the orientation of $\bb$ in the shear
eigenvector frame. First, if $\bb$ lies principally along $s_0$,
fieldlines are neither being stretched nor compressed, only
interchanged, and therefore they cascade without amplification. A
profile of $\bb$ in a plane perpendicular to $s_0$ has an appearance
similar to a passive scalar.  In this extreme alignment, the power
spectrum is $E_b(k) \, \sim \, k^{-1}$.

The other extreme is for the fieldlines to be aligned with $s_+$, so
that they are amplified as they cascade. If each reduction in scale is
accompanied by a self-similar amplification of energy, the resulting
power spectrum index is $E_b(k) \, \sim \, k^0$. The geometry of
fieldlines in the saturated state must be bracketed by these extremes,
therefore the subviscous magnetic spectrum will have an index between
$-1$ and $0$.  This is observed in figure \figfinalstate, where the
index is closer to $-1$. The alignment of $\bb$ in the eigenvector
frame is found to be preferentially along $s_0$, which also suggests
an index closer to $-1$.

We measure the saturated-state alignment of $\bb$ in the local shear
frame by evaluating $G_2=<(b_ib_jS_{ij})^2>,$ which is the variance of
the magnetic alignment. In the time-averaged saturated state, $G=0$,
and so the mean is zero. We compare this to the result obtained after
random-phasing $\bb$ (but not $\vv$). For a structured field, $G_2$
has one half the value as for a random-phased field. This establishes
that in the local shear frame, $\bb$ preferentially aligns with $s_0$.

\subsection{Forcing scale}

Simulation K4 is forced at $3\leq s \leq 4$, unlike the other
simulations which are forced at $1\leq s \leq 2$. Its purpose is to
determine if magnetic energy can occupy modes larger than the forcing
scale. The result from figure \figcji\, is that they do not.  The
initial conditions are taken from the saturated state of simulation A4
at t=19. The viscosity and resistivity are the same as for simulation
A4.

\hbox to \hsize{ \hfill \epsfxsize7cm \epsffile{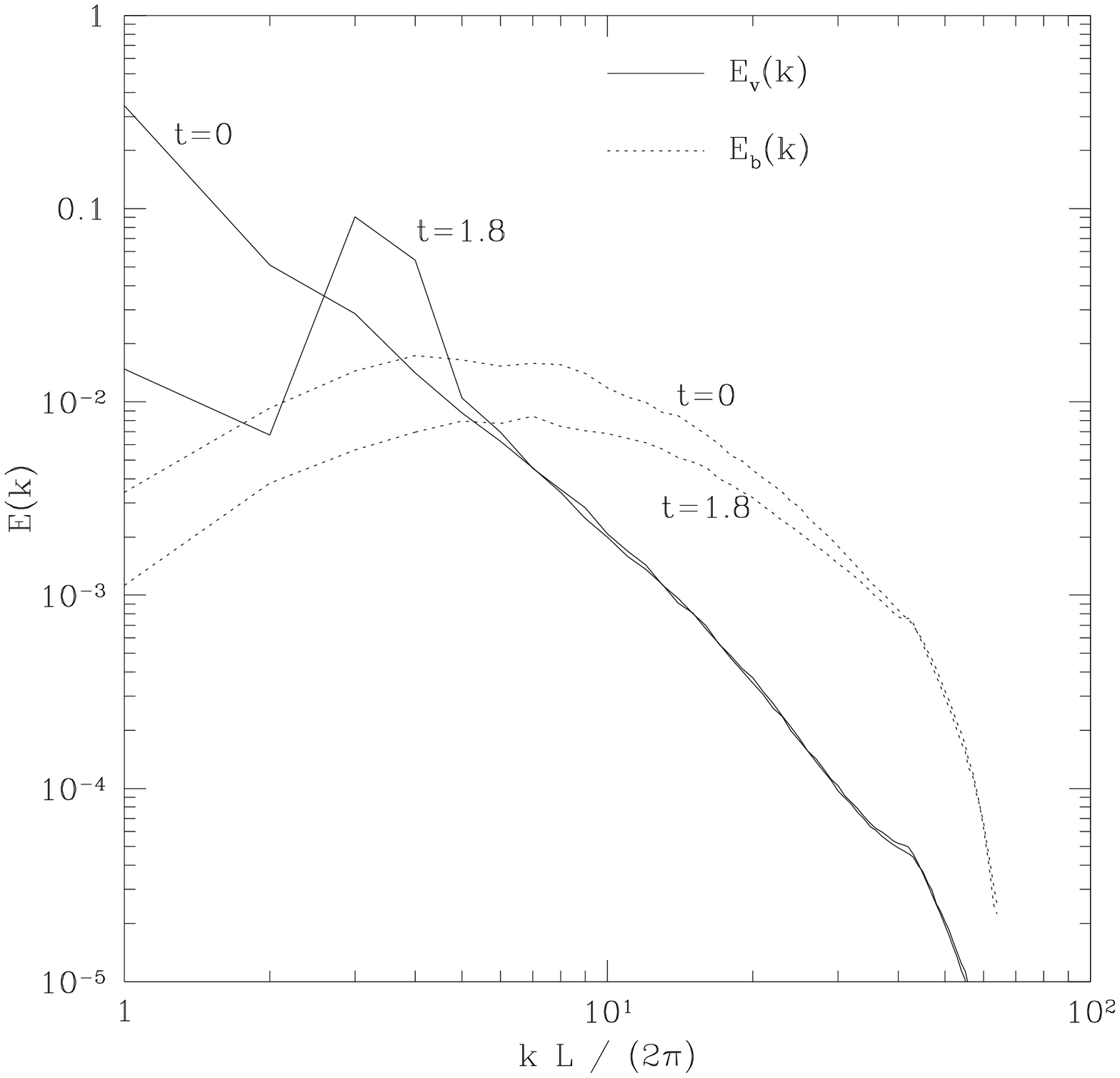} \hfill}
\vspace{-4mm} \noindent {\it Figure \figcji: In simulation K4,
magnetic energy does not occupy to a significant degree the
wavenumbers smaller than the forcing wavenumbers at $s=3,4$.}

\subsection{Turbulent diffusivity}

For $P_r < 1$, sub-resistive velocities hinder magnetic growth. We
establish this with two simulations having the same $\lambda_\eta$ and
different $\lambda_\nu$ (figure \figsubprandtla).  The first (S1) has
$\lambda_\nu = \lambda_\eta$, and the second (S2) has $\lambda_\nu
\sim 0.4 \lambda_\eta$, or $P_r = 0.4$. The second has
sub-resistive kinetic energy while the first does not. The resolution
of S2 is sufficient to fully capture the dynamics. The magnetic
energy in simulation S1 grows more quickly than that for simulation
S2.  The interpretation is that the sub-resistive eddies act as a
turbulent diffusivity on the magnetic field. We will see elsewhere
that turbulent diffusivity does not apply for $Pr \geq 1$.

In figure \figsubprandtlb, we observe that the magnetic field decays
for Prandtl numbers $0.2$ and $0.1$. However, it is possible that for
a simulation with these Prandtl numbers, and a larger grid, that the
field might grow instead.

\hbox to \hsize{ \hfill \epsfxsize7cm \epsffile{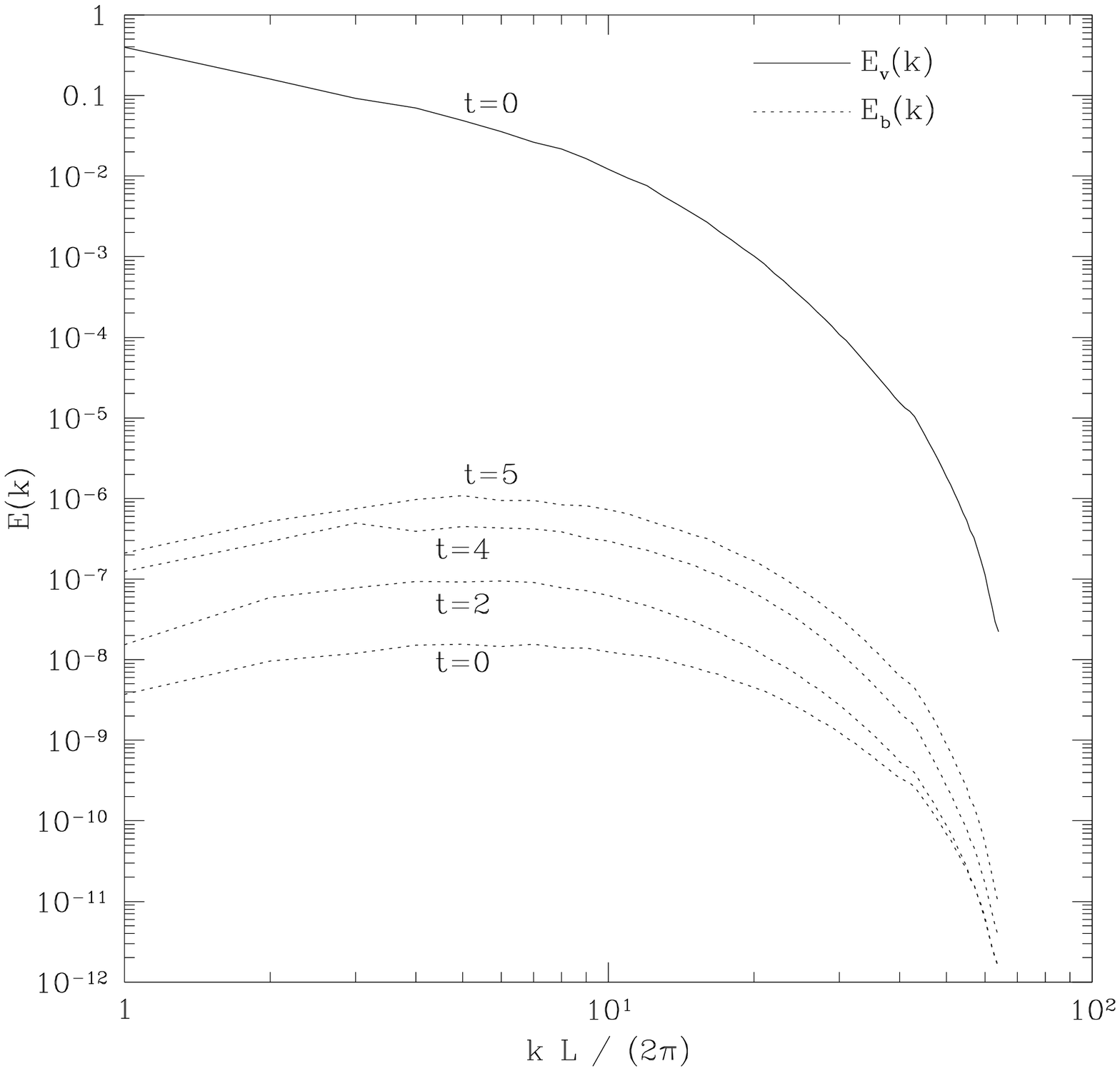} \hfill
\epsfxsize7cm \epsffile{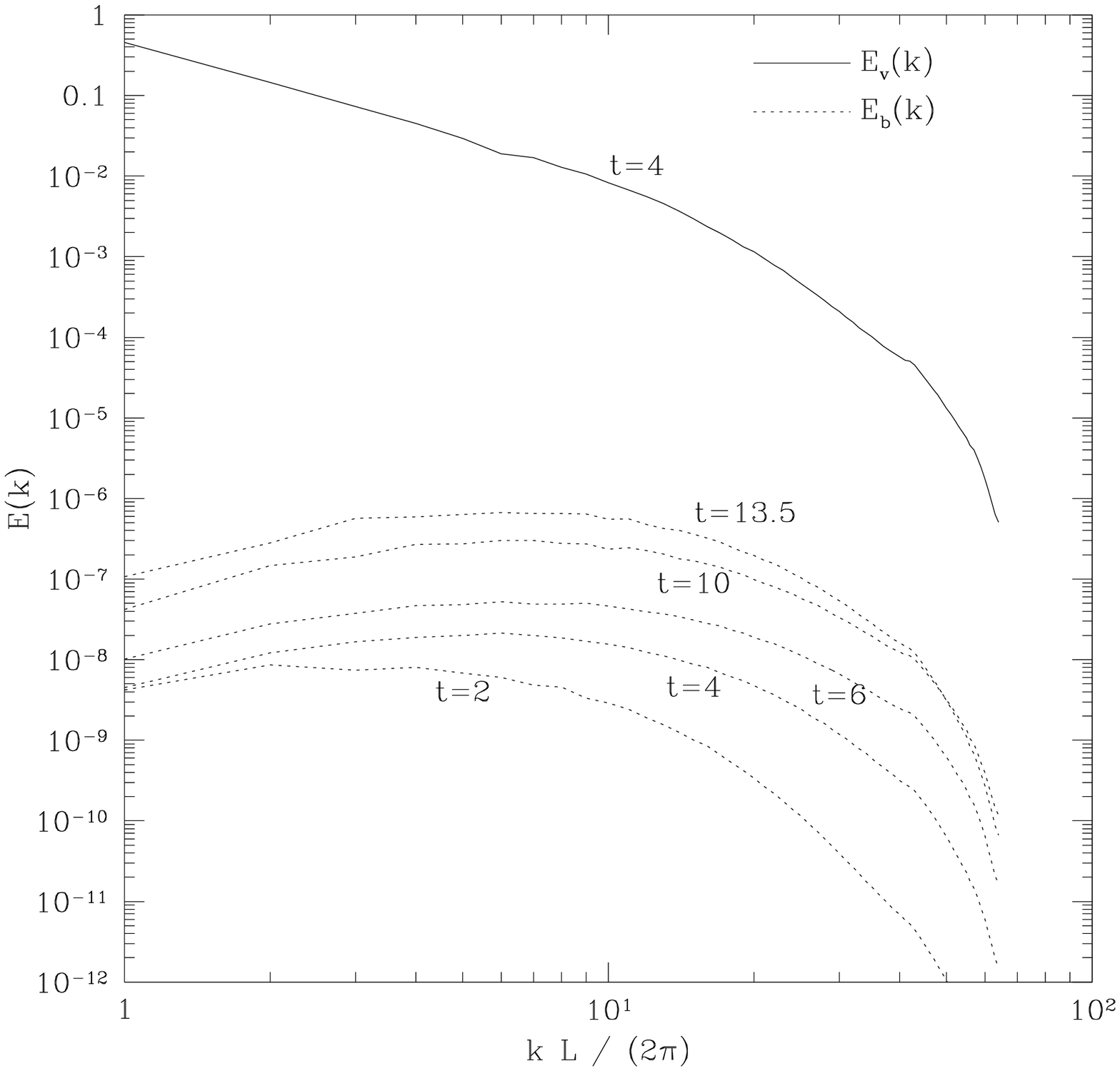} \hfill}

\vspace{-4mm} \noindent {\it Figure \figsubprandtla.  Left: simulation
S1, $\nu=10^{-3}$, $\eta=10^{-3}$, $Pr=1$.  The magnetic field grows
robustly.  Right: simulation S2, $\nu=4\cdot10^{-4}$,
$\eta=10^{-3}$, $Pr=0.4$.  The magnetic field grows slowly.}
\vspace{-1mm}

\hbox to \hsize{ \hfill \epsfxsize7cm \epsffile{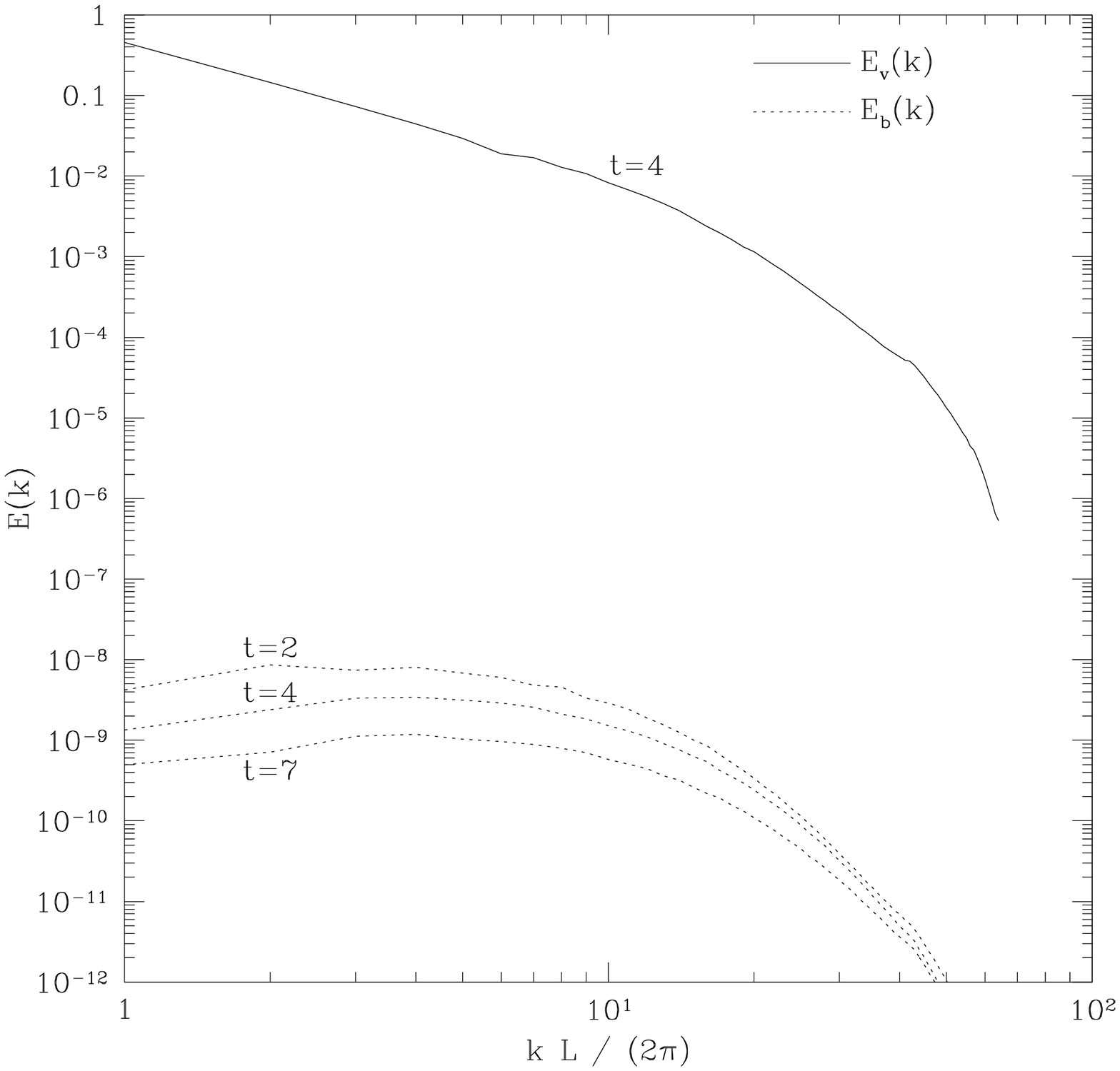} \hfill
\epsfxsize7cm \epsffile{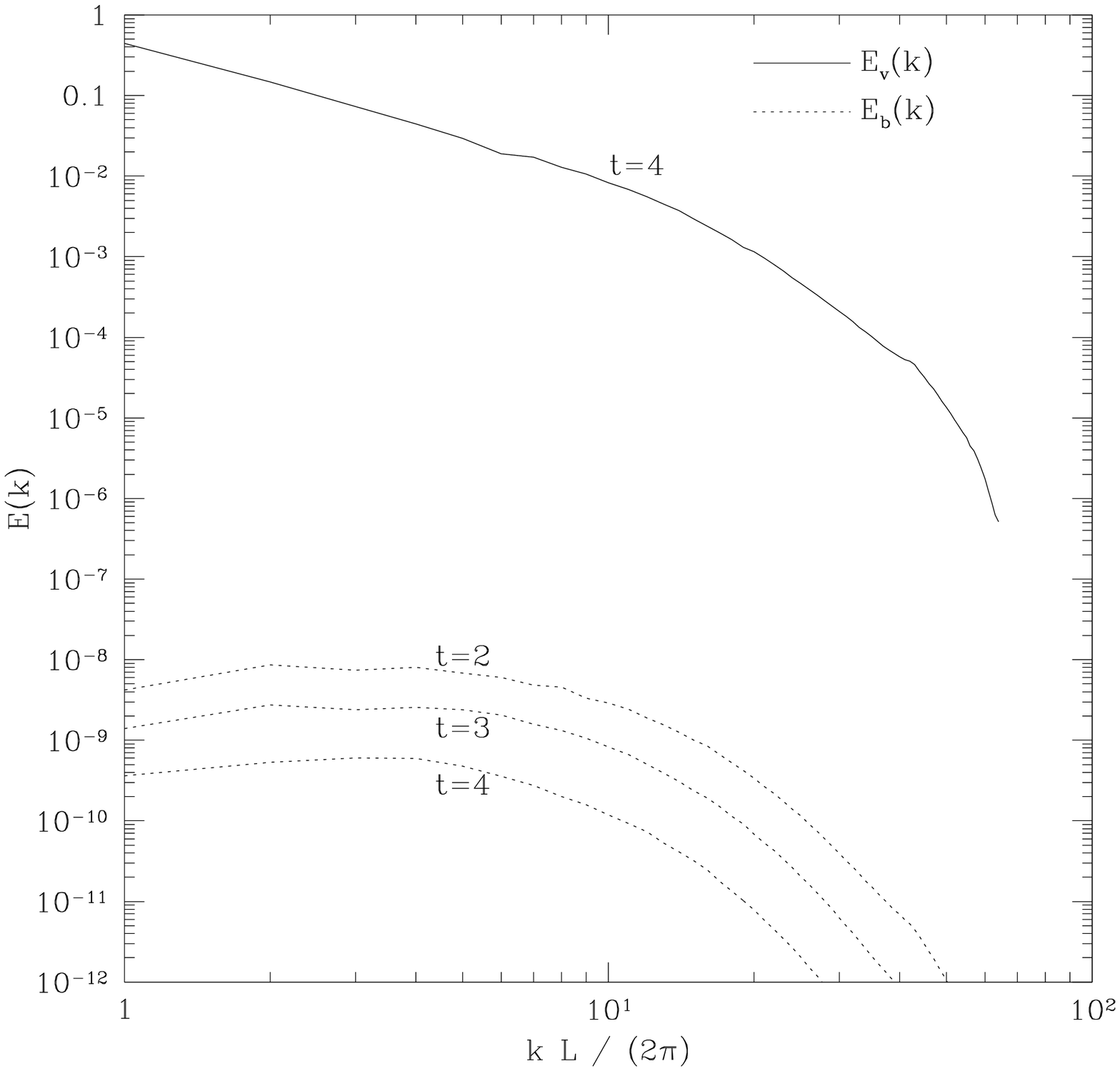} \hfill}

\vspace{-4mm} \noindent {\it Figure \figsubprandtlb.  Left: simulation
S3, $\nu=4\cdot10^{-4}$, $\eta=2\cdot10^{-3}$, $Pr=0.2$.  Right:
simulation S4, $\nu=4\cdot10^{-4}$, $\eta=4\cdot10^{-3}$,
$P_r=0.1$. In each of these simulations, the magnetic field decays.}
\vspace{-4mm}

\subsection{Pitfalls}

Subtle dangers threaten the validity of turbulent dynamo simulations.

For saturated dynamics, magnetic unwinding causes kinetic motions and
viscous dissipation at scales below $\lambda_\nu$ (section
Z). Therefore, the use of a hyperviscous operator in the Navier-Stokes
equation of the form $\nu_n \nabla^{2n} v$, where $n>1$, is not valid.
If used, it would anomalously destroy small-scale magnetic fields,
giving the false impression that the saturated magnetic spectrum does not
extend beyond the viscous wavenumber.

In hydrodynamic turbulence, inertial range dynamics are insensitive to
whether energy is removed at the inner scale by physical diffusivity
or by dealiasing. In MHD turbulence, dealiasing destroys folded
fieldline structure, increasing the unwinding and viscous dissipation
by a factor of $(\lambda_\parallel/\lambda_\perp)^2$, which can be as
large as $200$ in our simulations. Therefore, the resistivity must be
large enough to prevent magnetic energy from reaching the aliasing
scale.

The standard $\alpha-\Omega$ helical dynamo theory assumes that
locally, the uniform magnetic field component outweighs the
fluctuating component. It is of interest to know if the turbulent
dynamo yields a magnetic configuration in accord with this
assumption. We therefore emphasize zero helicity and high Prandtl
number in our simulations.  The astrophysical context has an inertial
range of $\sim 4$ decades and a Prandtl number of $\sim 10^{15}$.

The magnetic field backreacts on the turbulence as it increases in
energy after the linear regime, slowing down the shear timescale and
increasing the resistive scale according to $t_s \sim
\lambda_\eta^2/\eta$.  Although the magnetic scale increases, this
should not be interpreted as the formation of a large-scale field, as
the field is still at the resistive scale. Large resolution and
Prandtl number are required to yield a saturated state where the
magnetic energy scale is well separated from the kinetic energy scale.

The $\alpha-\Omega$ helical dynamo theory invokes a concept of
turbulent diffusivity acting on the magnetic field. In this picture,
turbulence with RMS velocity $v$ at scale $\lambda$ acts diffusively
on the magnetic field with an equivalent viscosity of $v\lambda$. This
is at odds with our results, which suggest instead that the magnetic
field is cascaded non-diffusively, and only by velocities which have
more energy than the magnetic field.

\section{Braginskii viscosity}

Simulations with Braginskii viscosity are qualitatively similar those
with Laplacian viscosity. In figure \figbraginski, we see that the
field grows quickly in the linear regime, and then reaches a saturated
state that is essentially the same as that for the Laplacian viscosity
simulations. The difference in the dynamics is that the Braginskii
viscosity does not oppose fieldline unwinding, and so the Alfven
scaling would apply for the unwinding time. We speculate that
topological constraints inhibit the unwinding of small-scale magnetic
fields. This is similar to our interpretation in section
\ref{largek}\, for why a large energy, large-scale magnetic field
ultimately evolves to a small-scale field.

\hbox to \hsize{ \hfill \epsfxsize7cm \epsffile{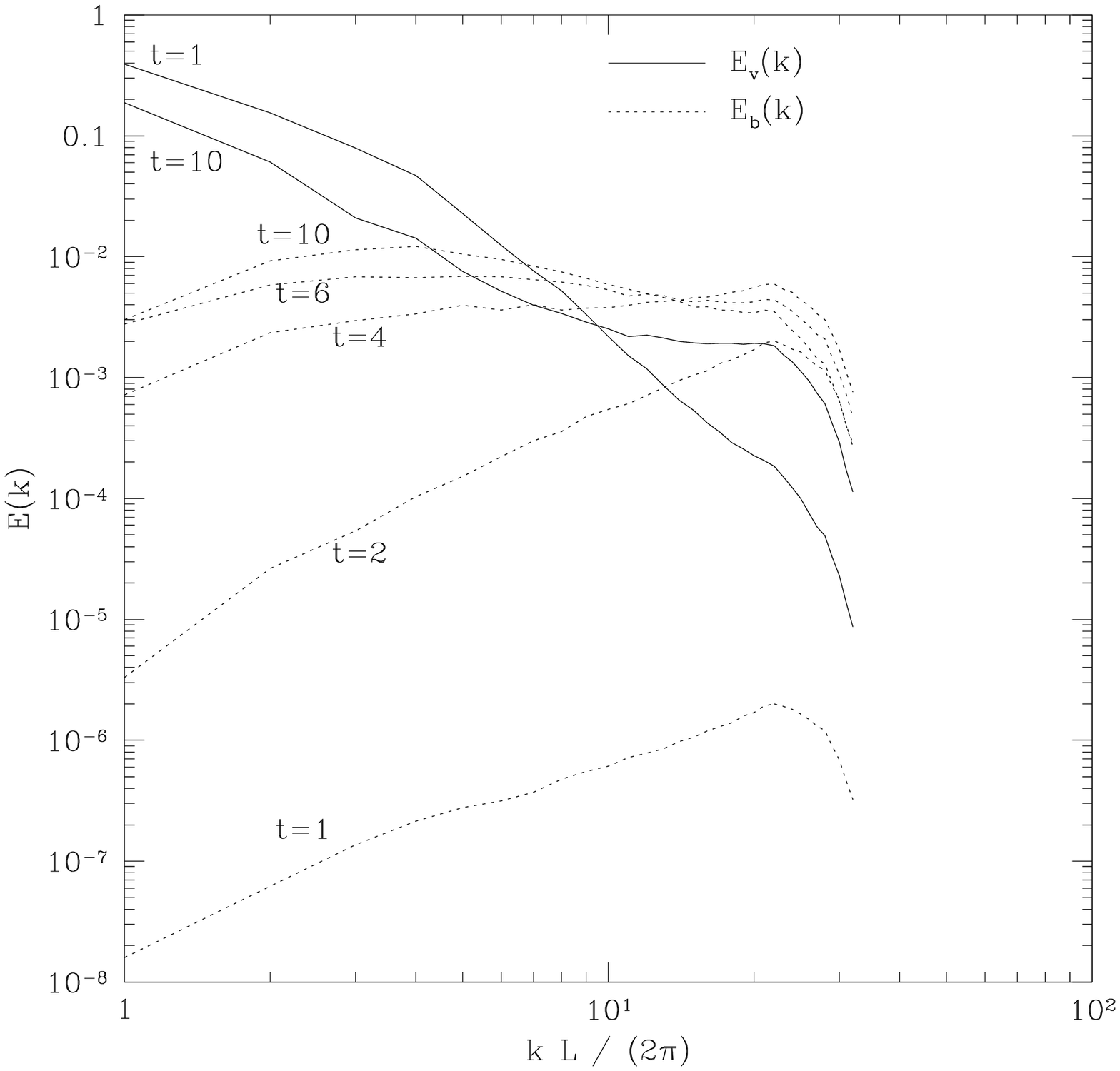} \hfill
\epsfxsize7cm \epsffile{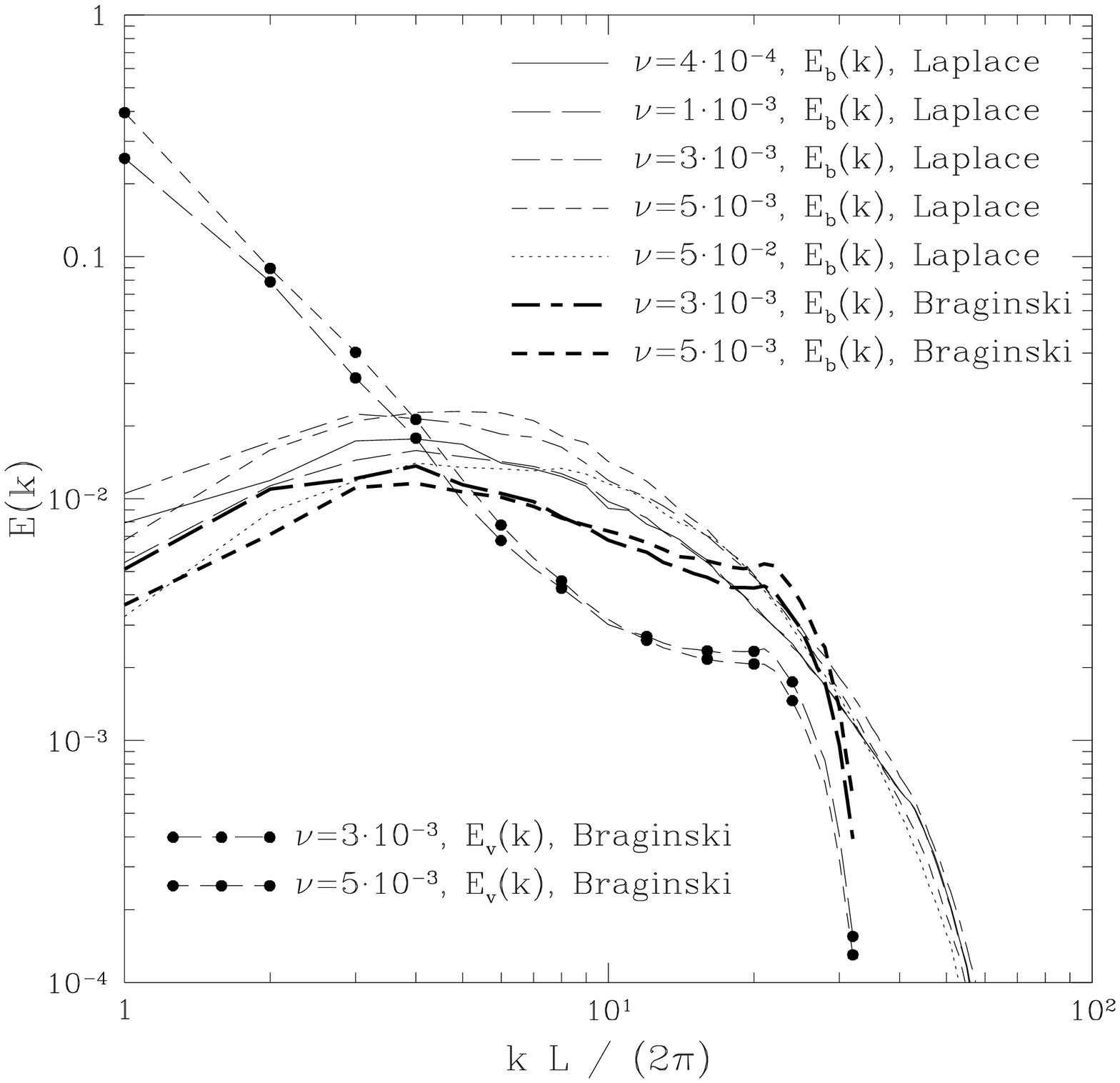} \hfill}

\vspace{-4mm} \noindent {\it Figure \figbraginski.  Left: The
evolution from the linear to the saturated state for the Braginsky
viscosity simulation T3, with $\nu=3\cdot10^{-3}$ and $\eta=10^{-4}$.
Right: The saturated magnetic spectra for two Braginskii viscosity
simulations (T2 and T3), shown together with the Laplacian viscosity
simulations from figure \figfinalstate. The resistivity is $10^{-4}$
in each. The magnetic spectra are nearly the same for each type of
viscosity.}  \vspace{-4mm}

\section{Acknowledgements}

We wish to thank James McWilliams, Alex Schekochihin, Benjamin
Chandran, Eric Blackman, Ellen Zweibel, Yoram Lithwick, and Axel
Brandenburg for helpful discussions.  We benefitted from the
supercomputers at the Caltech Center for Advanced Computing Resources
(10000 CPU hours) and at the National Center for Supercomputing
Applications at UIUC (10000 CPU hours), and from their very helpful
staff.

\appendix

\section{Index of simulations}

\begin{center} {\it Table \tablesima: Index of simulations} \end{center}
\small
\begin{center}
\begin{tabular}{cccccl}
\hline
ID  &  Grid & $\nu$           &$\eta$        &Pr& Notes\\
\hline
A0 &$128^3$&$5\cdot10^{-2}$&$0$            &    -&\\
A1 &$128^3$&$5\cdot10^{-2}$&$2\cdot10^{-5}$& 2500&\\
A2 &$128^3$&$5\cdot10^{-3}$&$1\cdot10^{-4}$&   50&\\
A3 &$128^3$&$3\cdot10^{-3}$&$1\cdot10^{-4}$&   30&\\
A4 &$128^3$&$1\cdot10^{-3}$&$1\cdot10^{-4}$&   10&\\
A5 &$128^3$&$4\cdot10^{-4}$&$4\cdot10^{-4}$&    1&\\
\hline
B1 &$256^3$&$5\cdot10^{-2}$&$1\cdot10^{-5}$& 2500&\\
B2 &$256^3$&$5\cdot10^{-3}$&$4\cdot10^{-5}$&  125&\\
B3 &$256^3$&$3\cdot10^{-3}$&$4\cdot10^{-5}$&   75&\\
B4 &$256^3$&$1\cdot10^{-3}$&$4\cdot10^{-5}$&   25&\\
B5 &$256^3$&$4\cdot10^{-4}$&$1\cdot10^{-4}$&    4&\\
B6 &$256^3$&$1\cdot10^{-4}$&$1\cdot10^{-4}$&    1&\\
\hline
L3 &$128^3$&$3\cdot10^{-3}$&$1\cdot10^{-4}$&    1&Init $\bb$ at $k=4$ with
  large energy.\\
L4 &$128^3$&$1\cdot10^{-4}$&$1\cdot10^{-4}$&    1&Init $\bb$ at $k=1$ with
  large energy.\\
\hline
T2 &$ 64^3$&$5\cdot10^{-3}$&$1\cdot10^{-4}$&   50&Braginskii viscosity\\
T3 &$ 64^3$&$3\cdot10^{-3}$&$1\cdot10^{-4}$&   30&Braginskii viscosity\\
T3b&$ 64^3$&$3\cdot10^{-3}$&$4\cdot10^{-4}$&  7.5&Braginskii viscosity\\
\hline
U4 &$128^3$&$1\cdot10^{-3}*$&$1\cdot10^{-5}$&  100&Erase $\vv$ from A4 and
  continue.\\
U4r&$128^3$&$1\cdot10^{-3}*$&$1\cdot10^{-5}$&  100&Like U4, \& with
                                                   random-phased \bb\\
\hline
K4 &$128^3$&$1\cdot10^{-3}$&$1\cdot10^{-4}$&  100&Force at s=3 \& 4.\\
\hline
A4w&$128^3$&$1\cdot10^{-3}$&$1\cdot10^{-6}$& 1000&Start from weak field\\
A5w&$128^3$&$4\cdot10^{-4}$&$1\cdot10^{-6}$& 1000&Start from weak field\\
A4s&$128^3$&$1\cdot10^{-3}$&$1\cdot10^{-6}$& 1000&Start from saturated state of A4\\
A5s&$128^3$&$4\cdot10^{-4}$&$1\cdot10^{-6}$& 1000&Start from saturated state of A5\\
\hline
S1 &$128^3$&$1\cdot10^{-3}$&$1\cdot10^{-3}$&    1&\\
S2 &$128^3$&$4\cdot10^{-4}$&$1\cdot10^{-3}$&  0.4&\\
S3 &$128^3$&$4\cdot10^{-4}$&$2\cdot10^{-3}$&  0.2&\\
S4 &$128^3$&$4\cdot10^{-4}$&$4\cdot10^{-3}$&  0.1&\\

\hline
\end{tabular}
\end{center}
\vspace{2mm}
\normalsize

\noindent The following properties are common to all simulations: The
box size is (1,1,1), and the kinetic energy is forced with a power of
unity. Forcing occurs within a sphere of radius 2 in Fourier lattice
space, except for simulation K4, which is forced at a radius of 3 and 4.

\section{Code} \label{code}

We evolve the incompressible MHD equations in Fourier space where they
take the form (Lesieur 1990)
\begin{eqnarray}
&
\partial_t \tilde{v}_\alpha
= -i k_\gamma \left( \delta_{\alpha\beta} - \frac{k_\alpha k_\beta}{k^2}
\right)
  \left( \widetilde{v_\beta v}_\gamma - \widetilde{b_\beta b}_\gamma \right) -
\nu k^{2n} \tilde{v}_\alpha& \label{eq:Fmomentum}, \\
&\partial_t \tilde{b}_\alpha = -i k_\beta
(\widetilde{v_\beta b}_\alpha - \widetilde{b_\beta v}_\alpha) - \eta k^{2n}
\tilde{b}_\alpha, &
\label{eq:Finduction}\\
& k_\alpha\tilde{v}_\alpha=0, \hspace{15mm}
  k_\alpha\tilde{b}_\alpha=0, &\label{eq:Fdivb} \\
&\partial_t \tilde{c} = -i k_\beta \widetilde{v_\beta c} - \nu_c
k^{2n} \tilde{c}. \label{eq:Fpassive}&
\end{eqnarray}


Equations \ref{eq:Fmomentum}, \ref{eq:Finduction}, and
\ref{eq:Fpassive} constitute a system of ordinary differential
equations with time as the dependent variable and the Fourier
coefficients $\{\tilde{v}_\alpha,\, \tilde{b}_\alpha,\, \tilde{c}\}$
as the independent variables. We employ a modified version of the
second order Runge-Kutta algorithm (RK2) to advance the variables in
time.

We make one departure from RK2 and treat diffusive terms with an
integrating factor. Consider an equation of the form
\be
\partial_t \tilde{q}(k) = A - \nu_n k^{2n} \tilde{q}(k),
\label{eq:modeleq}
\ee
where $A$ comprises the non-diffusive terms.  Its solution, with $A$
constant throughout the interval $\Delta t$, is
\be
\tilde{q}(\Delta t) = \left[ \tilde{q}(0) + \frac{A}{\nu_n k^{2n}}
(e^{\nu_n k^{2n}\Delta t} - 1) \right] e^{-\nu_n k^{2n}\Delta t}
\label{eq:modifiedE1}
\ee
We use this expression in place of E1 in each stage of RK2. To
lowest order in $\nu_n k^{2n}\Delta t$, equation \ref{eq:modifiedE1}
reduces to E1. However, it has the advantage that it yields stable solutions
to equation \ref{eq:modeleq} with constant $A$ for arbitrary values of
$\nu_n k^{2n}\Delta t$ whereas E1 yields unstable solutions for
$\nu_n k^{2n}\Delta t>2$

We employ standard dealiasing according to the $2/3$ rule (Canuto 1988)
except with the Braginskii and ambipolar terms, which require special
treatment (section \ref{braginsky}).

The code, written by Maron, is exhaustively discussed in Maron \&
Goldreich 2000.

Define the one-dimensional kinetic and magnetic energy spectra as
\begin{equation}
E_v = \int E_v(k) d\!k \hspace{15mm} E_b = \int E_b(k) d\!k
\end{equation}

\subsection{The Braginskii term}
\label{braginsky}


Bilinear terms in the MHD equations are calculated by
transforming the individual fields to real space, carrying out the
appropriate multiplications there, and then transforming the products
back to Fourier space.
This requires $N_1N_2N_3$ operations using the Fast
Fourier Transform (FFT) algorithm; $(N_1N_2N_3)^2$
operations would be needed to carry out the equivalent convolution
in Fourier space.

This economy comes at the price of either a 1/3 reduction in
resolution or an aliasing error. To appreciate this, consider the 1D product
\begin{eqnarray}
\widetilde{pq}(s) & = \frac{1}{N} \sum_{l} \left[
\sum_{s^\prime} \tilde{p}(s^\prime) e^{-2\pi is^\prime l/N}
\sum_{s^{\prime\prime}} \tilde{q}(s^{\prime\prime})
e^{-2\pi is^{\prime\prime} l/N}
\right] e^{2\pi isl/N} \nonumber \\
& = \frac{1}{N}
\sum_{s^\prime}\sum_{s^{\prime\prime}}
\tilde{p}(\ss^\prime)\tilde{q}(s^{\prime\prime})
e^{2\pi i (s^\prime + s^{\prime\prime}) l/N}
\delta_{s, s^\prime + s^{\prime\prime} + mN},
\end{eqnarray}
where $m$ is any integer. The $m=0$ terms comprise the convolution, and the
remainder the aliasing error.  To avoid the aliasing error, we set all
Fourier components with $|s| > N/3$ to zero both before we compute the real
space fields and again after we return the bilinear terms to Fourier space.
Truncation ensures that Fourier components of bilinear terms
with $m\ne 0$ vanish. Its cost is the reduction of the effective spatial
resolution from $N$ to $2N/3$.

The Braginskii term contains a product of 5 fields inside the gradient,
plus two divisions by $b^2$. When M terms are multiplied, dealiasing
should be applied for $|s| < N/(M+1)$, however the division will
introduce aliased terms regardless of the location of the
truncation. The appropriate procedure then becomes a matter of
engineering. The rule we apply to decide if the computation is
sufficiently accurate is that it should agree with the equivalent
computation performed on an infinitely large grid. We find that a
truncation of $|s| < N/3$ is inadequate for the Braginskii term, and
that a truncation of $|s| < N/6$ is more than adequate (figure
\figalias).

To test the properties of the dealiasing truncation, we evaluate the
Braginskii term for two grid sizes, with the larger grid regarded as
the "correct" result and the smaller grid the "test" result.  We apply
the same dealiasing truncation to both grids before and after the
computation before comparing. We then varied the location of the
dealiasing truncation for different pairs of grids. In table
\tablealias, pairs A-B, C-D, and E-F all share the same truncation.
The input fields are taken from simulation Z65 at $t=4$. The magnetic
field is dominated by small scale structure and presents a worst case
scenario for alias error.  In figure \figalias, we selected the
Braginskii term's $\hat{x}$ component and plotted it as a function of
$x$ at fixed $y$ and $z$ for each grid pair.  Test grid F, with a
truncation at $N/3$, shows poor agreement with the larger grid E.

We conclude that the MHD terms should be dealiased at $N/3$ and the
Braginskii term at $N/6$. We implement this by computing the MHD terms
on a size $N^3$ and the Braginskii term on a $(2N)^3$ grid. The
Braginskii result is then returned to a $N^3$ grid and added to
the MHD result. Aside from economizing resolution, this configuration
also economizes timestep. The timestep will be set by the $N^3$ grid,
which is twice as large as the timestep that would be necessary for the
$(2N)^3$ grid.

\begin{center} {\it Table \tablealias: Braginskii viscosity test
simulations.} \end{center}
\begin{center}
\begin{tabular}{lllllll}
\hline
ID & Grid & Truncation & \hspace{2cm} & ID & Grid & Truncation \\
\hline
A  & $64^3$ & $|s| < 32/6$ && B  & $32^3$ & $|s| < 32/6$ \\
C  & $64^3$ & $|s| < 32/4$ && D  & $32^3$ & $|s| < 32/4$ \\
E  & $64^3$ & $|s| < 32/3$ && F  & $32^3$ & $|s| < 32/3$ \\
\hline
\end{tabular}
\end{center}

The ambipolar term is computed by first obtaining the magnetic force
and then putting it in the place of $v$ in the induction equation.
This requres that we dealias the magnetic force before computing the
rest of the induction term.

\hbox to \hsize{ \hfill \epsfxsize7cm \epsffile{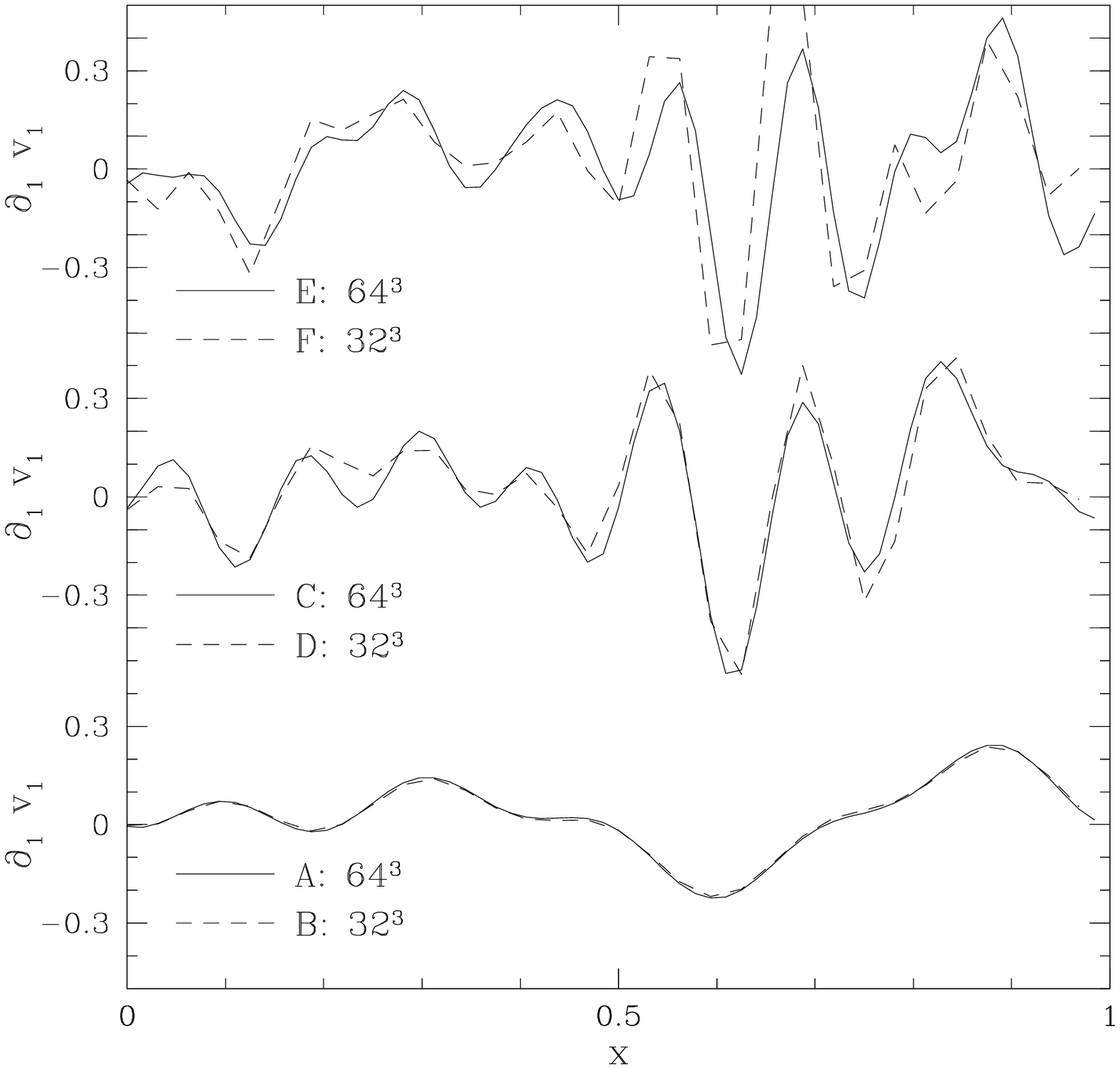} \hfill}
\vspace{-4mm} \noindent {\it Fig \figalias: Alias error in the
Braginskii term.  Curve B is the Braginskii term computed with a
dealiasing truncation at N/6, which agrees will with the equivalent
simulation on a larger grid.  The dealiasing truncation in curves D
and F is at N/4 and N/3 respectively.  The N/3 truncation leads to
poor agreement.}



\section{Random phase transformation} \label{randomphase}

Structure in the dynamical fields can be studied through comparison
with their random-phase realizations. The random phase transformation
is designed to remove the coherent structure while preserving the
power spectrum. The transformation is accomplished by giving each
Fourier mode vector $\vv(k)$ and $\bb(k)$ a new random orientation
while preserving the magnitude, subject to the constraint of
divergencelessness.


\begin{thebibliography}{}

\bibitem{arm95}
  Armstrong, J. W., Rickett, B. J., \& Spangler, S. R. 1995, ApJ, 443, 209
\bibitem{batch} Batchelor, G. K. 1950, Proc. Roy. Soc. Lon. A201, 405.
\bibitem{beck} Beck, R., Brandenburg, A., Moss, D., Shukurov, A. and Sokoloff,
  D. 1996, Annu. Rev. Astron. Astrophys. {\bf 34} 155.
\bibitem{bie50} Biermann, L., Z. Naturforsch 5a, 65(1950)
\bibitem{bra65} Braginskii, S. I. 1965, Rev. Plasma Phys. 1, 205.
\bibitem{cha97} Chandran, B., Cowley, S., 1998, PRL V80, no. 14.
\bibitem{cha00} Chandran, B.D.G., 2000, PRL, in press
\bibitem{che99} Chertkov, M., Falkovich, G.,  Kolokolov, I. \& Lebedev, V.
  1999, Phys. Rev. Lett. 83, 4065.
\bibitem[Glatzmaier and Roberts 95]{glaz} Glatzmaier, G.A. and Roberts,
  P.H., 1995, Nature, 377, 203.
\bibitem{fie99} Field, G.B., Blackman, E.G. \& Chou, H. 1999, Ap. J. 513, 638.
\bibitem{gne00} Gnedin N., Ferrara, A. and Zweibel, E.G. 2000, Ap. J. 539, 505.
\bibitem{gol95} Goldreich, P. \& Sridhar, S. 1995, ApJ, 438, 763
\bibitem{gol97} Goldreich P. \& Sridhar, S. 1997, ApJ, 485, 680
\bibitem{gru96} Gruzinov, A., Cowley, S., Sudan, R. 1996, Phys. Rev. Lett.
  77, 43, 42.
\bibitem{hei00} Heiles, C., Presented at the 4th Tetons summer conference:
  "Galactic structure, stars, and the interestellar medium."
\bibitem{kaz68} Kazantsev, A. P. 1968, Sov. Phys. JETP, 26, 1031.
\bibitem{kin98} Kinney, R. M., Chandran, B. D. G., Cowley, S. C., McWilliams,
  J. C., American Astronomical Society Meeting, 192, 66.18, 1998
\bibitem{kin00} Kinney, R. M., Chandran, B., Cowley, S., McWilliams, J. C.,
  2000, ApJ, 545, 907.
\bibitem{kra67} Kraichnan, R. \& Nagarajan, S. 1967, Phys. Fluids, 10, 859,
\bibitem[Kronberg (1994)]{kron} Kronberg, P. 1994, Rep. Prog. Phys 59 325.
\bibitem{kul92} Kulsrud, R., \& Anderson, S., 1992, ApJ, 396, 606
\bibitem{kul97} Kulsrud, R., Cen, R., Ostriker, J. P., \& Ryu D., 1997, ApJ,
  480, 481
\bibitem{kul69} Kulsrud, R., \& Pearce, W., 1969, ApJ, 156, 445
\bibitem{kul99} Kulsrud, R.M., 1999, Annu. Rev. Astron. Astrophys.
\bibitem{les90} Lesieur, M. 1990, Turbulence In Fluids (Dordrecht:Kluwer)
\bibitem{mar00} Maron, J., Goldreich, P., 2001, ApJ, 554, 1175
\bibitem{mar01b} Maron, J., Blackman, E., 2001, PRL, submitted.
\bibitem{mof78} Moffatt, H. K., 1978, Magnetic field generation in
  Electrically Conducting Fluids, Cambridge University Press: Cambridge
\bibitem{par79} Parker, E.N. 1979, Cosmical Magnetic Fields
  (Oxford:Oxford Univ. Press)
\bibitem{ros58} Rosenbluth, Kaufman, Phys Rev 109, 1, 1958
\bibitem{ruz88} Ruzmaikin, A. A., Shukurov, A. M., \& Sokoloff, D. D.,
  Magnetic Fields of Galaxies (Dordrecht: Kluwer)
\bibitem{sch01} Schekochihin, A. A., Boldyrev, S. A., Kulsrud, R. M. 2001,
  astro-ph/0103333, submitted to ApJ.
\bibitem{sch00} Schekochihin, A., Ph.D. Thesis, Princeton University, 2000.
\bibitem{sch01} Schekochihin, A., Cowley, S., Maron, J., Malyshkin, L.,
  PRE, accepted.
\bibitem{spa98} Spangler, S. R. \& Cordes, J., ApJ, 505, 766, 1998
\bibitem{spi98} Spitzer, ``Physical Processes in the Interstellar Medium", 1978
\bibitem{sri94} Sridhar S. \& Goldreich, P. 1994, ApJ, 432, 612
\bibitem{taylor} Taylor, G. B., Allen, S. W., Fabian, A. C., 1999, Proceedings
  of the workshop: ``Diffuse thermal and relativistic plasma in galaxy
  clusters", 77-81.
\bibitem{vai86} Vainshtein, S. I., \& Kichatinov, L. L. 1986, J. Fluid Mech.,
  168, 73
\bibitem{vallee} Vallee, J. P., Fundamental Cosmic Physics, 1998, 19, 319-422.
\bibitem{zwe97} Zweibel, E.G. and Heiles, C. 1997, Nature, 131.




\end{thebibliography}
\end{document}